\documentclass[preprint, review,3p,times, 10pt]{elsarticle}
\usepackage{cite}
\usepackage{amsmath,amssymb,amsfonts}
\usepackage{algorithmic}
\usepackage{graphicx}
\usepackage{textcomp}
\usepackage{comment}
\usepackage{footnote}
\usepackage{threeparttable}
\usepackage{romannum}
\usepackage{longtable}
\usepackage{lscape}
\usepackage{soul}
\usepackage{array, makecell}
\usepackage{float}
\usepackage{multirow}
\usepackage{multicol}
\usepackage{tabularx}
\usepackage{pifont}
\usepackage{subcaption}
\usepackage{booktabs}
\usepackage[hyphens]{url}
\usepackage{hyperref}
\hypersetup{colorlinks=true,breaklinks=true}

\journal{arxiv}

\begin{document}

\begin{frontmatter}
\title{Self-DenseMobileNet: A Robust Framework for Lung Nodule Classification using Self-ONN and Stacking-based Meta-Classifier}

\author[1]{Md. Sohanur Rahman}\ead{mdsohanur2087@gmail.com}
\author[2]{Muhammad E. H. Chowdhury\corref{correspondingauthor1}}\ead{mchowdhury@qu.edu.qa}
\author[1]{Hasib Ryan Rahman}\ead{hasibryan99@gmail.com}
\author[1]{Mosabber Uddin Ahmed\corref{correspondingauthor1}}\ead{mosabber.ahmed@du.ac.bd}
\author[3]{Muhammad Ashad Kabir\corref{correspondingauthor1}}\ead{akabir@csu.edu.au}
\author[4]{Sanjiban Sekhar Roy}\ead{sanjibanroy09@gmail.com}
\author[1]{Rusab Sarmun}\ead{rusabsarmun@gmail.com}
\cortext[correspondingauthor1]{Corresponding authors.}
% \cortext[correspondingauthor1]{Corresponding author: Department of Electrical and Electronics Engineering, University of Dhaka, Dhaka 1000, Bangladesh.}
% \cortext[correspondingauthor2]{Corresponding author: Department of Electrical Engineering, Qatar University, Doha 2713, Qatar.}
% \cortext[correspondingauthor3]{Corresponding author: School of Computing, Mathematics and Engineering, Charles Sturt University, Bathurst, NSW 2795, Australia.}

\affiliation[1]{organization={Department of Electrical and Electronic Engineering, University of Dhaka}, city={Dhaka}, postcode={1000}, country={Bangladesh}}
\affiliation[2]{organization={Department of Electrical Engineering, Qatar University}, city={Doha}, postcode={2713}, country={Qatar}}
\affiliation[3]{organization={School of Computing, Mathematics and Engineering, Charles Sturt University}, city={Bathurst}, state={NSW}, postcode={2795}, country={Australia}}
\affiliation[4]{organization={School of Computer Science and Engineering, Vellore Institute of Technology}, city={Vellore}, state={Tamil Nadu}, country={India}}
% \affiliation[4]{organization={Department of Electrical Engineering, Qatar University}, city={Doha}, postcode={2713}, country={Qatar}}

\begin{abstract}
In this study, we propose a novel and robust framework, Self-DenseMobileNet, designed to enhance the classification of nodules and non-nodules in chest radiographs (CXRs). Our approach integrates advanced image standardization and enhancement techniques to optimize the input quality, thereby improving classification accuracy. To enhance predictive accuracy and leverage the strengths of multiple models, the prediction probabilities from Self-DenseMobileNet were transformed into tabular data and used to train eight classical machine learning (ML) models; the top three performers were then combined via a stacking algorithm, creating a robust meta-classifier that integrates their collective insights for superior classification performance. To enhance the interpretability of our results, we employed a class activation mapping algorithm such as ScoreCAM to visualize the decision-making process of the best-performing model.
Our proposed framework demonstrated remarkable performance on internal validation data, achieving an accuracy of 99.28\% using a Meta-Random Forest Classifier. When tested on an external dataset, the framework maintained strong generalizability with an accuracy of 89.40\%. These results highlight a significant improvement in the classification of CXRs with lung nodules.

\end{abstract}

\end{frontmatter}

\section{Introduction}
\label{sec:introduction}
Lung cancer is a leading cause of cancer-related mortality worldwide, affecting both males and females. Over 1.7 million people are affected with lung cancer annually~\citep{Thandra2021,Bray2018}. A key challenge in diagnosing lung cancer is its tendency to remain asymptomatic until the disease reaches advanced stages. This delayed onset of symptoms significantly hampers effective treatment as the patient's condition becomes more precarious~\citep{WHO2023}. Early discovery of lung cancer can contribute notably as it can scale down the mortality rates. The prevalence of lung nodules in the lungs is notably high in the adult population, although less than 5\% of these nodules are malignant~\citep{ATM24877}. A benign solitary lung nodule may arise due to a residual scar from a previous infection caused by fungal, tuberculous, bacterial, or parasitic agents. Alternatively, it may represent a persistent infection, though this is less common~\citep{Walter2021}. To determine the malignancy of a nodule, a biopsy may be necessary. However, this procedure carries inherent risks, including pneumothorax, hemorrhage, air embolism, and tumor seeding \citep{wu2011diagnosis}. 
 
The main diagnostic imaging modality for thoracic disorders such as lung cancer, tuberculosis, and pneumonia is the CXR \citep{Hooda2018, Malik2023}. This diagnostic method is known for its affordability, availability, and ability to detect pathogenic changes. Radiologists often struggle to diagnose early-stage lung cancer via CXR due to the small size of lung nodules \citep{Juan2023}. Additionally, nodules may be obscured by nearby organs like the heart or ribs, increasing the complexity of detection. Even skilled radiologists may miss cancerous nodules when examining numerous CXRs \citep{White1999}. 

% This is where CAD frameworks become valuable, as they 
Artificial intelligence (AI) based computer-aided diagnosis (CAD) frameworks can classify these nodules, enhancing the efficacy of their discovery \citep{Schalekamp2014}. AI has the potential to serve as a valuable adjunct in the classification of lung nodules, offering an additional viewpoint that is both reliable and consistent. Considering the accessibility of CXR screening in rural regions \citep{Ausawalaithong2018}, the utility of AI for classifying nodules in CXRs is anticipated to provide significant implications in the timely identification of lung nodules. Radiologists can employ CAD to precisely classify different samples and later efficiently look for nodules in nodule samples without wasting time looking into healthy samples. 

Deep learning (DL) (a branch of AI) provides a tremendous edge in developing biomedical-based challenges spanning from distinguishing and detecting diseases to the development of biomedical tools. Pulmonary diseases like tuberculosis can easily be detected using DL techniques~\citep{Rahman2020}. Moreover, transfer learning algorithms are proven to be reliable in distinguishing and localizing pneumonia from CXRs, which can assist radiologists in identifying infected samples and delivering appropriate service to patients \citep{Rahman2020}. Additionally, during the COVID-19 pandemic, DL techniques developed prompt localization pipelines to classify between infected patients and healthy patients, as well as structured early warning systems for superior service in healthcare facilities, which possibly could lessen burdens on healthcare providers \citep{Tahir2021, Degerli2021, Rahman2022}.

The utilization of AI-based CAD frameworks in identifying lung nodules from CXRs has been extensively studied \citep{Ausawalaithong2018, Girvin2012, Juan2023, Schultheiss2020}. However, existing research~\citep{Juan2023,Girvin2012} has primarily concentrated on the design and efficacy of CAD systems, overlooking the impact of image pre-processing techniques and the sustained performance of AI models. In addition, some existing studies~\citep{Ausawalaithong2018,Schultheiss2020} leverage pre-trained models that incorporate numerous deep learning layers. However, to address the complexities associated with existing approaches, we developed a new framework specifically designed for the classification of CXRs, including lung nodules of varying sizes. Our framework introduces a novel deep learning architecture, Self-DenseMobileNet, which enhances model diversity and adaptability while maintaining a streamlined and efficient design. Inspired by recent advances in self-organized operational neural networks (SelfONNs), this approach addresses the limitations of traditional convolutional neural networks (CNNs) by introducing higher levels of network heterogeneity and optimizing internal operations \citep{s23167156}. The integration of SelfONN layers allows the model to dynamically adjust its operations during training, leading to a more adaptive and efficient feature extraction process. The architecture leverages bottleneck residual blocks with self-organized operations to improve network diversity without increasing computational overhead. The key contributions of this study are outlined below:

\begin{itemize}
    \item We have proposed the Self-DenseMobileNet architecture, featuring a shallow architecture optimized through iterative refinement with various enhanced image samples. 

    \item We have designed a stacking-based meta-classifier, leveraging three best-performing classical machine learning models (out of eight ML models) to enhance classification performance.

    \item We compared our model's performance with state-of-the-art AI models, including DenseNet201, MobileViTv2-0.50, MobileViTv2-0.75, and ResNet152. We validated our model on completely unseen images as part of our external validation to demonstrate the model's accuracy and reliability on entirely new data.

    \item We used a class activation mapping algorithm such as ScoreCAM to visually illustrate the interpretability of our model's predictions. The generated heatmaps indicate regions of interest that likely contain nodules.
\end{itemize}

% The paper is organized as follows. Section \ref{sec:methodology} provides an overview of the methodology employed, including information on the datasets used, the pre-processing steps undertaken, and the approaches employed for image enhancement and augmentation. Section \ref{sec:experiment} describes the procedures used for training the model, whereas Section \ref{sec:results} presents the results and analysis of the nodule classification. Finally, Section \ref{sec:conclusion} concludes the paper.

\section{Related works}
\label{sec:relatedwork}
In the domain of lung nodule classification, numerous datasets and methodologies have been developed over the years. These techniques, which are summarized in Table \ref{tab:medical_image_analysis}, range from image enhancement and data augmentation to more sophisticated methods like deep learning, classical machine learning, stacking, and model interpretability. An exploration of the key contributions in each of these areas is presented below.

Early lung image analysis studies have heavily focused on image enhancement to improve the clarity and diagnostic value of CXRs. Various techniques, including background removal, contrast enhancement, and lung field segmentation, were employed to improve the visibility of critical regions such as lung nodules, aiding more accurate diagnosis \citep{lo1995artificial}. To further enhance the image features, biologically inspired filters like the Laplacian of Gaussian (LoG) and Gabor kernels were applied, helping to sharpen edges and highlight key features in the lung area \citep{coppini2003neural}. Another technique, normalize cross-correlation (NCC), was used to generate nodule-like pattern-enhanced images by employing rib templates, which distinguished nodules from surrounding rib structures \citep{chen2020pulmonary}. Additionally, image intensity was adjusted by cropping to the first and 95th percentiles, followed by intensity normalization to the range [0,1], ensuring uniform brightness and contrast levels across the dataset \citep{Chatterjee2024Feb}. 

% The use of a square structuring element with an 8-connected neighborhood ensured that the operation could cleanly separate the lung regions from the background, resulting in smoother masks \citep{Moura2022Jan}. To reduce noise further, a Gaussian smoothing filter was applied to the X-ray images, enhancing overall image quality by blurring unnecessary details. Following this, a logarithm operator, also known as the pixel logarithm operator, was used to improve local contrast, particularly by enhancing low-intensity pixel values that could otherwise obscure important features like nodules \citep{shaheed2023computer}. 

As the this domain further progressed, data augmentation techniques emerged to tackle one of the primary challenges in medical imaging: the limited availability of large, labeled datasets. By artificially expanding datasets, these augmentation methods significantly improved the generalizability and robustness of ML models, allowing them to perform better on unseen data. Among the commonly used techniques were horizontal inversion and angle rotation, which introduced variability into the training data. Images were randomly flipped horizontally 50\% of the time and rotated by 90°, 180°, and 270°, thereby increasing the diversity of the dataset and helping models learn to recognize nodules from multiple angles and perspectives \citep{chen2020pulmonary, Gozes2019Jul, Lyu}. 

% In one particular approach, benign nodules were rotated 90 degrees counterclockwise to artificially increase the number of training samples. These augmented images were then incorporated into the dataset, resulting in a total of 10,752 benign nodule samples. To maintain a balanced dataset, half of the malignant nodule samples were randomly removed, reducing the count to 11,249 malignant nodules. This balancing ensured that the three classes (benign nodules, malignant nodules, and non-nodules) were nearly in a 1:1:1 ratio, which is critical for minimizing bias in model predictions. Ultimately, this process yielded a dataset of 33,001 nodules in total, providing a well-rounded foundation for training robust models \citep{Lyu}.

In recent years, DL has emerged as the dominant approach for handling complex tasks in medical image analysis, particularly in lung nodule detection. Early studies focused on simpler architectures, such as 2D CNNs and ANNs, which were employed for detecting lung nodules in both CT scans and X-ray images. These initial models were applied to datasets such as the JSRT dataset, proving to be effective but limited in their ability to capture the depth and complexity of medical images. As the field evolved, more advanced architectures were developed, such as Residual U-Net, Multi-level CNNs, and CDCNet, along with the incorporation of 3D ANNs. These models offered a deeper understanding of image patterns, improving the detection and segmentation of lung nodules by capturing 3D spatial relationships. This advancement proved particularly beneficial for analyzing lung fields in CXR and CT scans, including cases associated with diseases like COVID-19. The combination of deep learning techniques with various preprocessing methods, such as noise reduction and image enhancement, has allowed for even more precise segmentation of lung fields and accurate detection of nodules. These preprocessing steps help models focus on the critical features within an image, further improving diagnostic accuracy. By leveraging the power of these deep learning architectures, the field has seen a significant leap in the ability to reliably segment and detect lung abnormalities, contributing to higher overall accuracy and robustness in medical image analysis \citep{lan2018run, Li2023Feb, Hung2023Sep, Lyu, coppini2003neural, lo1995artificial, Shi2020, shaheed2023computer, Chatterjee2024Feb}.

Although DL has become the dominant approach in recent years, classical ML methods—such as Random Forest—continue to play a crucial role in lung image analysis. These techniques have proven particularly valuable in situations where model interpretability and computational efficiency are priorities. For instance, Random Forest image classification using radiomics features has been highly effective in extracting meaningful patterns from lung images, enhancing both diagnostic accuracy and transparency in decision-making. Moreover, hybrid models that combine the strengths of CNNs with Random Forest have demonstrated significant success. These models leverage the powerful feature extraction capabilities of CNNs while utilizing Random Forest for robust classification, resulting in improved performance in clinical settings. Such combinations offer a balance between the complexity of deep learning and the simplicity of classical machine learning, providing both high accuracy and interpretability—a critical requirement in healthcare applications \citep{Moura2022Jan, shaheed2023computer}.

The use of model stacking, although less common than other techniques, has shown considerable promise in recent studies focusing on lung image analysis. Stacking involves combining multiple models to enhance performance metrics by leveraging the strengths of different algorithms. This ensemble method creates a more robust predictive system by mitigating the weaknesses of individual models. For instance, Vision Transformer (ViT)-based models have been employed for deep learning feature extraction due to their capability to capture global image features effectively. These deep features provide rich representations of the lung images, which can improve classification tasks. Subsequently, Random Forests trained with a variety of parameters have been stacked using a voting-based algorithm to further improve accuracy. This approach integrates the predictions of multiple Random Forest classifiers, allowing the ensemble to make more reliable decisions based on the majority vote. While stacking has not been widely adopted across all studies, its potential to significantly boost overall performance is noteworthy. The ability to combine different algorithms and harness their individual strengths presents a valuable opportunity to enhance diagnostic accuracy in medical imaging. Therefore, further exploration of stacking techniques is warranted and could lead to substantial advancements in future lung image analysis research \citep{shaheed2023computer}. 

One of the key challenges in deploying machine learning models in healthcare is ensuring their interpretability. While early studies primarily focused on improving performance, they often overlooked the importance of making these models transparent and understandable for clinical use. However, more recent works have shifted towards developing models that are not only accurate but also interpretable, especially in the analysis of CT scans. Several techniques have been introduced to address this need for transparency. Methods such as occlusion, saliency mapping, input X gradient, guided backpropagation, integrated gradients, and DeepLIFT have been used to make deep learning models more interpretable by highlighting the features that contribute most to the model’s predictions \citep{Chatterjee2024Feb}. In some approaches, the predicted output is combined with Gradient-weighted Class Activation Mapping (Grad-CAM), which provides visual explanations for the model's predictions by identifying regions in the input image that are most relevant to the decision-making process. This further enhances interpretability, making it easier for clinicians to trust and understand the model’s decisions \citep{shaheed2023computer}. In the context of CT scan analysis, a 3D interpretable hierarchical semantic convolutional neural network (HSNet) has been employed to recognize specific features of lung nodules, including calcification, margin, texture, sphericity, and malignancy. By extracting distinct features for each semantic label, the HSNet model significantly improved AUC values for each classification task, demonstrating its effectiveness in providing more transparent predictions \citep{Hung2023Sep}. Additionally, interpretability has been further enhanced using the SHAP-RFECV approach, which evaluates the most important features and their impact on the model’s predictions. In this case, XGBoost was used as the classifier, offering valuable insights into the decision-making process and ensuring that the model remains transparent and understandable, which is crucial in a healthcare setting \citep{Moura2022Jan}.

\begin{table}[ht]
\centering
\caption{Summary of state-of-the-art studies on lung image analysis}
\label{tab:medical_image_analysis}
% \begin{tabular}{p{2.5cm} p{2cm} p{2cm} p{2cm} p{2cm} p{2cm} p{2cm}}
\begin{tabular}{lccccccc}
\toprule
\shortstack{Study} & \shortstack{Image \\ Enhancement} & \shortstack{Augmen-\\tation} & \shortstack{Deep \\ Learning} & \shortstack{Classical \\ Machine \\ Learning} & \shortstack{Stacking} & \shortstack{Interpret-\\ability} \\
\midrule
\citet{lo1995artificial} & \ding{51} & \ding{55} & \ding{51} & \ding{55} & \ding{55} & \ding{55} \\
\citet{coppini2003neural} & \ding{51} & \ding{55} & \ding{51} & \ding{55} & \ding{55} & \ding{55} \\
\citet{chen2020pulmonary} & \ding{51} & \ding{51} & \ding{51} & \ding{55} & \ding{55} & \ding{55} \\
\citet{Gozes2019Jul} & \ding{55} & \ding{51} & \ding{51} & \ding{55} & \ding{55} & \ding{55} \\
\citet{lan2018run} & \ding{51} & \ding{55} & \ding{51} & \ding{55} & \ding{55} & \ding{55} \\
\citet{Lyu} & \ding{55} & \ding{51} & \ding{51} & \ding{55} & \ding{55} & \ding{55} \\
\citet{Hung2023Sep} & \ding{55} & \ding{55} & \ding{51} & \ding{55} & \ding{55} & \ding{51} \\
\citet{Chatterjee2024Feb} & \ding{51} & \ding{55} & \ding{51} & \ding{55} & \ding{55} & \ding{51} \\
\citet{Moura2022Jan} & \ding{51} & \ding{55} & \ding{55} & \ding{51} & \ding{55} & \ding{51} \\
\citet{shaheed2023computer} & \ding{51} & \ding{55} & \ding{51} & \ding{51} & \ding{51} & \ding{51} \\
\hline
This Study & \ding{51} & \ding{51} & \ding{51} & \ding{51} & \ding{51} & \ding{51} \\
\bottomrule
\end{tabular}
\end{table}

In this study, a comprehensive approach is adopted, integrating multiple methodologies to classify lung nodules effectively. The methods include image enhancement techniques to improve the quality of input images, data augmentation to address the limitations of dataset size, and advanced deep learning architectures to achieve high-performance diagnostics. Unlike earlier studies, this work combines classical machine learning approaches for a more balanced analysis, ensuring both accuracy and simplicity. Model stacking is employed for the final classification, with a strong focus on interpretability, aligning this research with the current trends in explainable AI for healthcare applications. The progression from basic neural networks to sophisticated deep learning models, coupled with the increasing emphasis on data augmentation, stacking, and interpretability, highlights the continuous advancements in lung nodule classification. These collective works provide a solid foundation for further innovations, ultimately contributing to the development of more accurate, reliable, and trustworthy diagnostic tools.

\section{Methodology}
\label{sec:methodology}

\begin{figure}[!t]
    \centering
    \includegraphics[width=1\textwidth]{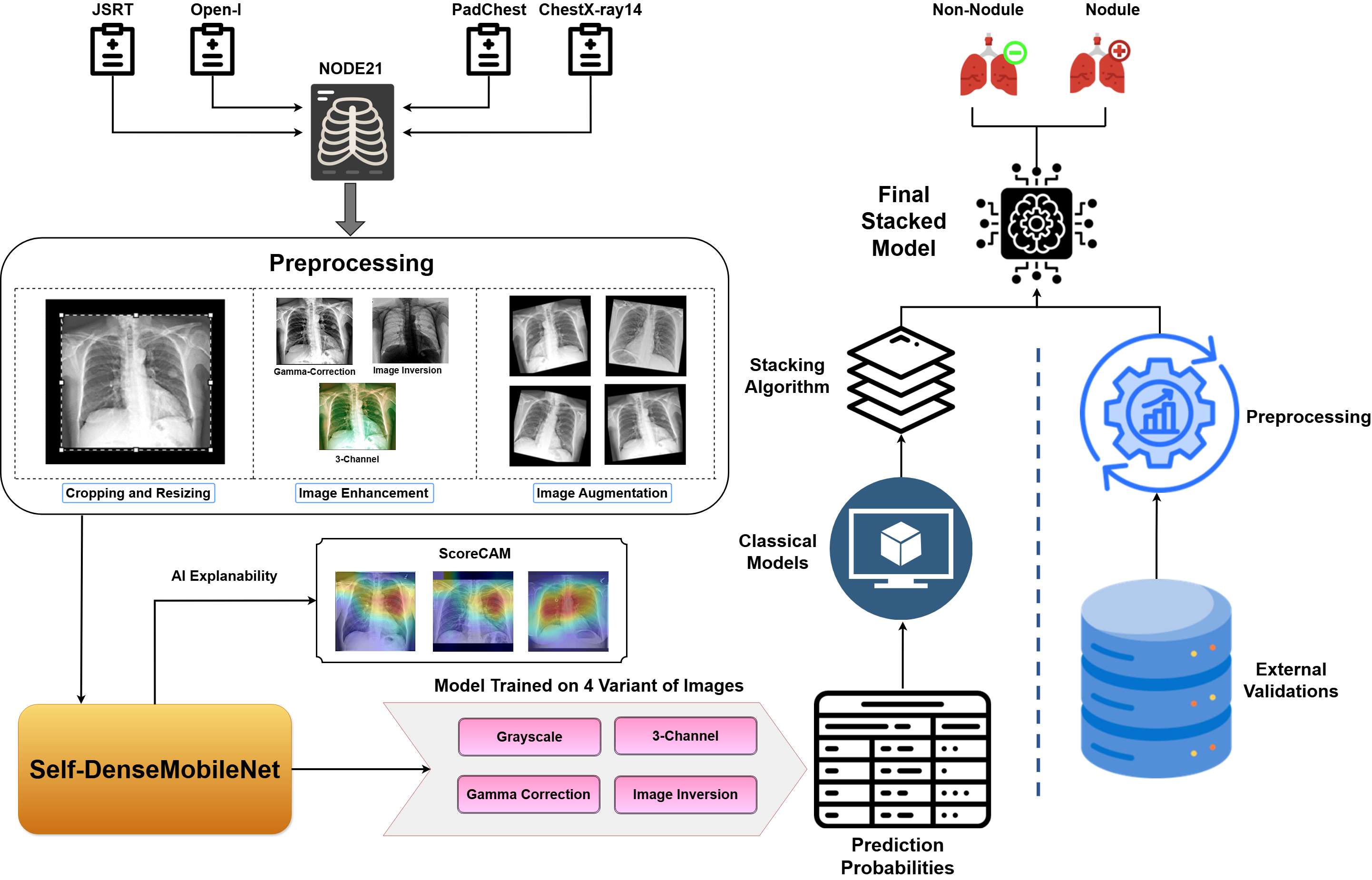}
    \caption{Methodological overview diagram for nodule classification.}
    \label{fig:methodology}
\end{figure}

Figure~\ref{fig:methodology} depicts an overview of our methodology classifying lung nodule. The process begins with clipping and resizing to remove unnecessary borders and standardize image dimensions (Section \ref{sec:preprocessing}). Various image enhancement techniques, such as gamma correction, image colour inversion, and three-channel preprocessing, were employed to improve image quality. These techniques utilized grayscale-converted samples, gamma-corrected images, and CLAHE-processed samples to enhance the dataset's representation (Section \ref{sec:enhancement}). To ensure balanced sample distribution across all classes, image augmentation methods including rotation, flipping, and cropping were applied (Section \ref{sec:augmentation}).
A comprehensive analysis of the dataset followed, evaluating the effectiveness of transfer learning methods. The training process was then conducted using our proposed shallow Self-DenseMobileNet model, which provided faster inference times while maintaining high performance (Section \ref{sec:selfdense}). For interpretability, the ScoreCAM algorithm was used to generate visual explanations of the model's predictions, making them more understandable for human review (Section \ref{sec:cam}). The predicted probabilities from four distinct versions of Self-DenseMobileNet, each trained on different enhanced image variants, were transformed into tabular format for further processing. Classical machine learning models were subsequently trained on this tabular data (Section \ref{sec:model_classification}), with the top-performing models combined into a stacking-based meta-classifier to generate the final results. Finally, external validation on an independent dataset was conducted to confirm the generalizability of the framework (Section \ref{sec:external}).

\subsection{Dataset Description}
\label{sec:datasetdescription}
We used the publicly available NODE21~\citep{10479589} dataset, which is sourced from several well-known publicly available datasets, including JSRT \citep{Shiraishi2012}, PadChest \citep{Bustos2020}, ChestX-ray14 \citep{Wang2017}, and Open-I \citep{Demner-Fushman2012}. The NODE21 dataset consists of 4882 images, of which 1134 images contain nodules, while the remaining 3748 images are categorized as non-nodule, as they do not exhibit any nodules. For the purpose of binary classification, images containing nodules were labeled as `1' or positive class, and non-nodules were labelled as `0' or negative class.

To ensure robust external validation, we curated an additional dataset by combining samples from two distinct sources \citep{Nguyen2022, Kermany2018}. The first source was the VinBigData CXR dataset \citep{Nguyen2022}, which includes 18,000 images encompassing 14 different forms of thoracic anomalies. In this dataset, lung nodules are categorized as ``Nodule/Mass" with the identified ID 8. From this dataset, we selected 500 CXRs with clearly defined nodules to ensure clear visual representation. For non-nodule samples, we sourced healthy CXRs from~\citep{Kermany2018} dataset, which contains 108,312 images, with 51,140 labeled as ``Normal". From this subset of ``Normal" images, we carefully selected 753 images that visually resembled non-nodule samples in NODE21. These images were used as the non-nodule samples in our external validation. Table~\ref{table:datasets} provides a summary of both the NODE21 dataset and the additional dataset used for external validation.

\begin{table}[!ht]
\centering
\caption{Distribution Summary of Nodule and Non-Nodule samples}
\begin{tabular}{lccc}
\toprule
Dataset & Nodule & Non-Nodule & Total \\ \midrule
NODE21~\citep{10479589} & 1134 & 3748 & 4882 \\ 
% \hline
External Validation \citep{Nguyen2022, Kermany2018} & 500 & 753 & 1253 \\ \bottomrule
\end{tabular}
\label{table:datasets}
\end{table}

\subsection{Preprocessing}
\label{sec:preprocessing}
To standardize the images, we performed several preprocessing steps. First, we removed uniform borders from the images as they typically contained minimal informative content. This step focused on retaining the essential visual elements of the images. Next, we applied energy-based normalization to equalize intensity values across the images, ensuring a uniform intensity representation, enhancing comparability across the dataset \citep{Philipsen2015}. The images were then cropped to isolate the region of interest, eliminating any extraneous data that could obscure relevant features. Finally, all images were resized to a fixed dimension of $1024\times1024$ pixels while maintaining the original aspect ratio. To achieve this uniform dimension, padding was applied to the shorter side of the images as needed. Our external dataset also underwent same sequence of preprocessing operations as the NODE21 images.

\subsection{Image Enhancement}
\label{sec:enhancement}
\begin{figure}[!ht]
    \centering
    \includegraphics[width=1\textwidth]{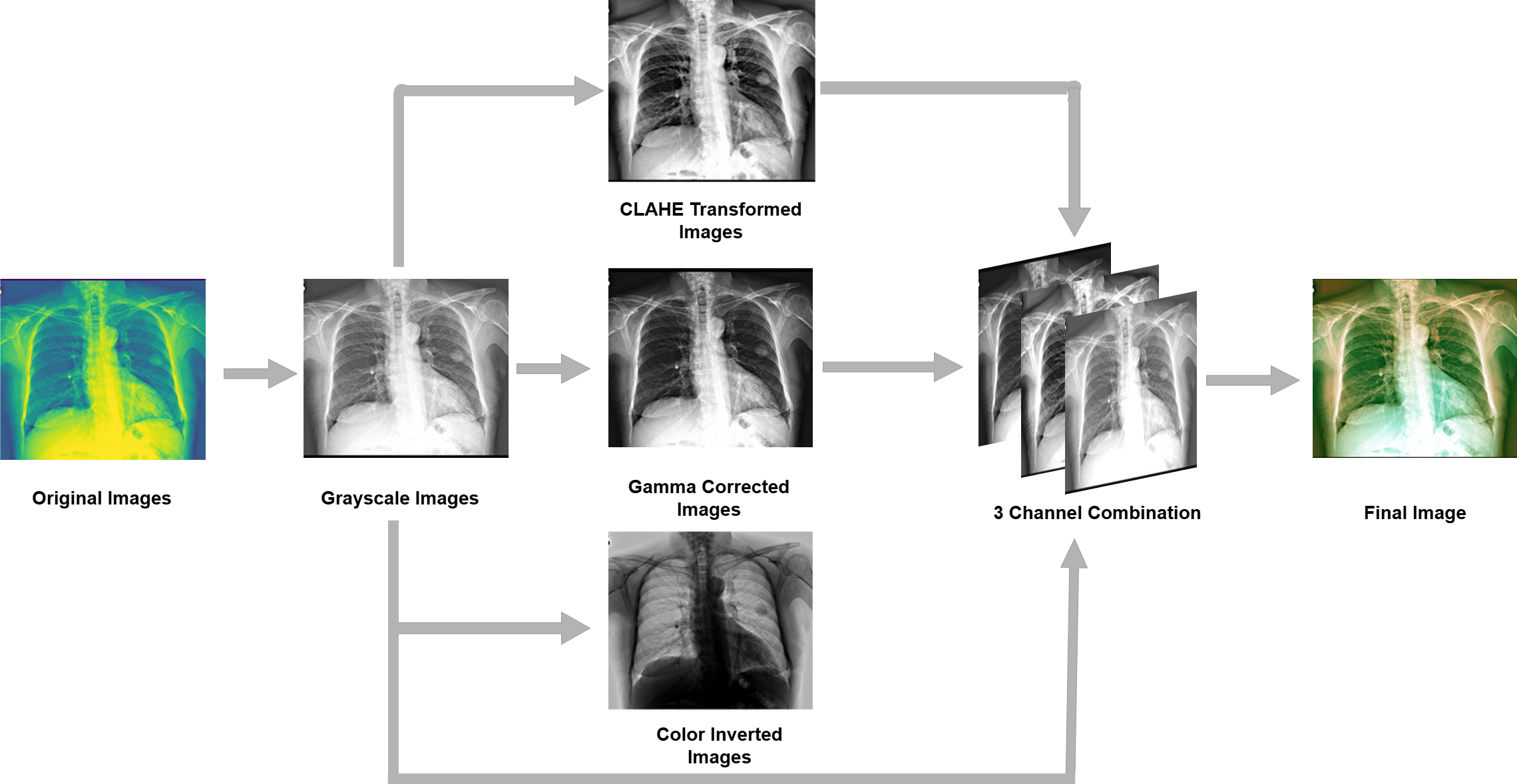}
    \caption{3-channel preprocessing using original and enhanced images.}
    \label{fig:channel-preprocessing}
\end{figure}

% The field of medical image processing offers significant benefits through its ability to provide comprehensive and non-invasive examinations of internal anatomical structures. In recent years, this technology has emerged as a crucial tool for advancing healthcare. Unlike conventional still photographs, 
Digital radiographs, such as CXRs, exhibit complex structures and diverse modalities which require specialized techniques to maintain data integrity and restore attenuated features during analysis and processing \citep{Salem2019}. A notable challenge in the observation of CXRs is the inadequate provision of optimal viewing conditions, which can impede the accurate interpretation of images. Enhancing the quality of medical images is crucial, as it helps radiologists in making critical decisions regarding patient care \citep{Salem2019}. As depicted in Figure~\ref{fig:channel-preprocessing} and discussed below, several widely used image enhancement techniques in the field of medical imaging~\citep{vidyasaraswathi2015review} were applied in this study to improve the visual clarity and diagnostic utility of the CXRs. 
% Histogram-based methods have demonstrated the potential to enhance image contrast and restore the original distribution of pixel intensities. In the context of X-ray imaging and related circumstances, when the objective is to capture image sequences or films, it is a common practice to utilize low-level exposure until the area of interest is recognized. In instances of this nature, it is imperative to augment the quality of real-time image samples \citep{Reza2004}. The contrast-limited adaptive histogram equalization (CLAHE) technique is utilized to enhance the contrast of an image by dispersing the intensities of pixels in a manner that avoids excessive amplification of noise and maintains the integrity of local features. Gamma correction is another technique employed to enhance pixel brightness and eliminate extraneous elements within an image \citep{dash2023}.

% Additional approaches that can be employed include color inversion and multi-channel combination, among others. When considering image enhancement techniques, it is important to consider the dimensions of the image and the range of pixels.

% \subsubsection{Grayscale}
% \label{sec:grayscale}
\paragraph{Grayscale} 
% The original images were provided in ‘.mha’ format. To facilitate the use of image enhancement algorithms, they were converted to ‘.png’ format. 
Images were converted to Grayscale, which reduces the image complexity to only black, white and gray pixels. This conversion significantly decrease the computational cost associated with model training, including reduced memory usage and processing time \citep{Bui2017}.

% \subsubsection{Color Inversion}
% \label{sec:inversion}
\paragraph{Image Colour inversion} Image Colour inversion is a valuable technique for highlighting significant regions within a radiograph by reducing glare and enhancing the visibility of key areas~\citep{Chowdhury2021}. In CXRs, inverting colours can improve the visibility of features or abnormalities. Darker regions, such as lung nodules, may become more pronounced against a lighter background, thus aiding AI model in identifying potential abnormalities. In this study, the colour inversion was applied to grayscale images, to enhance the contrast and visibility of critical features.

% \subsubsection{Gamma Correction}
% \label{sec:gamma}
\paragraph{Gamma Correction}
Gamma correction is a technique for adjusting an image's global luminance, making it particularly effective for enhancing images that are overexposed or too dark. This method allows for the manipulation of pixel intensity levels, thereby significantly improving the visual quality of images with low brightness or faded appearance. Gamma correction facilitates the process of extrapolating common characteristics from CXRs that exhibit excessively dark or bright colour patterns. The mathematical representation of gamma correction~\citep{Oommen2022} is provided below:
\[ \text{Output} = C \times \text{Input}^{\gamma} \tag{1}\]
Where, \( C \) and \( \gamma \) are constants. The input images are multiplied up with the gamma factor. This compensates for the non-uniform distribution of pixels in the image. In our case, \( C \) is 1 and \( \gamma \) is 2 as Gamma Correction were applied to Grayscale images. If the value of multiplier, \(C \), is greater than 1, the image becomes too dark, causing the presence of nodules become vague in CXRs. Conversely, if \(C \) is less than 1, the image becomes too bright, making the nodules blend into the background and disappear. After experimenting with different values of \( \gamma \), we found that a value of 2 provided the most optimal performance for our models. 

% \subsubsection{Multi-Channel Combination}
% \label{sec:multichannel}
\paragraph{Histogram-based Enhancement}
Histogram-based methods offer significant benefits in image augmentation, compression, and segmentation \citep{Pietka2000}. Histogram Equalization (HE) is a technique that equalizes the probability distribution of intensity levels across an image, resulting in a uniform distribution that enhances contrast and expands the dynamic range~\citet{das2015histogram}. Another effective approach is Contrast Limited Adaptive Histogram Equalization (CLAHE), which improves local contrast by redistributing pixel intensities dynamically~\citet{khan2020contrast}. Unlike HE, CLAHE addresses the issue of excessive noise amplification by limiting the contrast enhancement to localized regions within the image. This technique enhances image quality by adjusting the tile size and applying clipping, leading to significant improvements in local contrast and visual detail at the pixel level \citep{Joseph2017Jan}. In this study, CLAHE was applied using the Albumentations toolkit with a clip limit of 8.0 and a tile grid size of (4, 4). The clip limit of 8.0 controls the threshold for contrast clipping to prevent over-enhancement of noise, while the tile grid size of (4, 4) divides the image into smaller regions to apply localized contrast adjustments.

\paragraph{Channel merging}
% X-ray images are typically represented in grayscale, which consists of a single colour channel. By merging two distinct enhanced images (such as CLAHE and Gamma corrected) with the original grayscale image, a composite 3-channel image is created (depicted in Figure~\ref{fig:channel-preprocessing}), offering a more detailed and informative representation. This multichannel integration approach enhances the analysis of low-quality chest X-rays (CXRs) by providing richer information and improving interpretability. The use of multichannel structures in image processing has been shown to significantly boost performance in both ML and DL frameworks. Studies have demonstrated that incorporating multichannel data can increase accuracy and other performance metrics by approximately 5-6\% \citep{Tolstokulakov, Cai2022}. For instance, \citet{Nneji2022} introduced an innovative method for combining multiple polarization channels of an image, which resulted in a notable performance improvement in distinguishing pneumonia-affected CXRs from healthy ones, achieving an accuracy of up to 98.3\%. The conversion of grayscale images to multi-channel images enhances the amount of information available within a single representation. This augmentation of information enhances the accuracy of nodule classification in CXRs.

%X-ray images are typically represented in grayscale, which consists of a single colour channel. 
CXR images are typically represented using a single intensity channel, particularly in grayscale format.By merging two distinct enhanced images (such as CLAHE and Gamma corrected) with the original grayscale image, a composite 3-channel image is created (depicted in Figure~\ref{fig:channel-preprocessing}), offering a more detailed and informative representation. This multichannel integration approach enhances the analysis of low-quality chest X-rays (CXRs) by providing richer information and improving interpretability.
The use of multichannel structures in image processing has been shown to significantly boost performance in both ML and DL frameworks. Studies have demonstrated that incorporating multichannel data can increase accuracy and other performance metrics by approximately 5-6\% \citep{Tolstokulakov, Cai2022}. For instance, \citet{Nneji2022} introduced an innovative method for combining multiple polarization channels of an image, which resulted in a notable performance improvement in distinguishing pneumonia-affected CXRs from healthy ones, achieving an accuracy of up to 98.3\%. The conversion of grayscale images to multi-channel images enhances the amount of information available within a single representation. This augmentation of information enhances the accuracy of nodule classification in CXRs.

% The process of converting grayscale images to multi-channel combined images enhances the level of information contained within a single image. This augmentation of information subsequently contributes to the improved accuracy of ML models in the classification of nodules in chest X-rays.

% In this work, the initial steps involved the implementation of CLAHE and Gamma correction techniques. Then, the grayscale image was blended with it. 
% Figure \ref{fig:channel-preprocessing} displayed such integration and the final product of such preprocessing.

\subsection{Image Augmentation}
\label{sec:augmentation}
% In some scenarios where the available data is insufficient for effectively training classifying models, it becomes exceedingly important to increase the size of the dataset in the training set. The dataset used in this study comprised of 3,748 images, accounting for 76\% of the total, which did not exhibit any nodules. Conversely, 1,134 images, including 24\% of the dataset, did contain nodules. This leads to a class imbalance event, resulting in trained models exhibiting bias towards a singular class \citep{Goceri2023}. In the field of medical image augmentation, the often-employed techniques for enhancing CXR images include rotation, flipping, and scaling. \citet{Tandon2022} examined several fundamental augmentation techniques and observed a significant enhancement in the overall performance score. The metrics of accuracy, specificity, sensitivity, and F1-score all exhibited values exceeding 99\%. \citet{Shi2020} conducted an investigation on the utilization of Generative Adversarial Networks or GAN-based augmentation techniques for the purpose of enhancing Lung Nodule detection in their study. Image scaling and translation techniques are often employed in the context of augmentation. The utilization of techniques such as colour alteration, random erasing, or shearing was observed to have a detrimental impact on performance.
When a dataset exhibits class imbalance, augmenting the data becomes crucial, as such imbalance can lead to models exhibiting bias toward the more prevalent class \citep{Goceri2023}. Several image augmentation techniques are commonly employed to enhance CXR datasets. Techniques such as rotation, flipping, and scaling are frequently used to increase the diversity of the training data. For instance, \citet{Tandon2022} investigated these fundamental augmentation techniques and reported significant improvements in overall performance metrics. Additionally, \citet{Shi2020} investigated the use of Generative Adversarial Networks (GANs) for data augmentation to improve lung nodule detection. While their study demonstrated that GAN-based augmentation performed well for lung images, it was resource-intensive. Techniques like image scaling and translation proved to be more effective than other augmentation methods, whereas approachces such as colour alteration, randaom erasing, and shearing had a detrimental impact on performance.

In this study, the dataset comprised 3,748 images (76\% of the total) non-nodules and 1,134 images (24\% of the total) with nodules, resulting in a class imbalance with non-nodule images nearly three times more numerous than nodule images. To address this imbalance, augmentation techniques were applied to the training sets within each fold. These techniques included random rotation (both clockwise and counterclockwise) within an angle range of 0-15 degrees, cropping around the perimeter by 10\%, and horizontal flipping \citep{Shorten2019, Rahman2020}. These augmentation methods increased the number of nodule samples in the training set, bringing their quantity closer to that of the non-nodule samples.

% Due to a significant imbalance in distribution between the two classes- nodule and non-nodule, each fold of the dataset exhibited an uneven distribution of image samples. Specifically, the number of non-nodule samples was nearly three times more than the number of nodule samples. To address this imbalance, the training sets within each fold were augmented using techniques such as rotation (both clockwise and counterclockwise) by an angle range of 0-15 degrees, followed by cropping around the perimeter by 10\%, and finally horizontal flipping. All of these operations were performed randomly \citep{Shorten2019, Rahman2020}. As a result, this method increased the number of nodule samples in the training set close to the non-nodule samples.

\subsection{Nodule Classification}
\label{sec:noduleclassification}
We employed four state-of-the-art transfer learning techniques -- DenseNet201, MobileViTv2-0.50, MobileViTv2-0.75 and ResNet152, and our proposed Self-ONN based architecture, Self-DenseMobileNet, to classify between nodule and non-nodule images. To enhance the performance of our proposed model, we extracted prediction probabilities from four Self-DenseMobileNet models trained on four variants of images: grayscale, gamma correction, image colour inversion, and 3-channel images. These probabilites were compiled into a tabular data format with labels (1 for nodule and 0 non-nodule) and used to train eight classical ML models. Three of these base models, selected for their accuracy, were combined using a stacking algorithm to create the final meta-classifier.

\subsubsection{Transfer Learning}
\label{sec:transfer}
Transfer learning utilizes pre-trained model weights that have been trained on large datasets, thus eliminating the need to train a model from scratch \citep{Weiss2016}. This approach can identify patterns in input images with a high accuracy, resulting in robust and reliable outcomes. In this study, we used four pre-trained models -- DenseNet201, MobileViTv2-0.50, MobileViTv2-0.75, and ResNet152.

\paragraph{DenseNet201}
Densely-connected-convolutional network (DenseNet) is a neural network architecture that has direct connections between each layer and all subsequent layers, resulting in a feed-forward network architecture. DenseNet implements dense connections among layers. Every layer in the network receives input from their preceding layer and transmits its feature maps to all the succeeding layers. The high level of unity within the network allows the reuse of features and allows for the smooth flow of gradients. DenseNet exploits bottleneck layers with the objective to reduce computational expenses, which emerge as a result of dense connections \citep{Huang2017}. DenseNet-201 is a modified version of the initial DenseNet structure, featuring an increased number of layers, namely 201 layers \citep{Chowdhury2021}. The DenseNet201 model has demonstrated high performance in segmented lung images for Tuberculosis classification \citep{Rahman2020}.

\paragraph{MobileViT}
Designed to mix the strengths of CNNs with ViTs, MobileViT is a light-weight, mobile-friendly hybrid network \citep{mehta2022cvnets,mehta2022mobilevit}. Unlike traditional CNNs, these models employ transformers, originally devised for NLP tasks \citep{Dosovitskiy2020}, to process images. Incorporating detachable self-attention in place of the conventional multi-head attention (MHA) in the transformer blocks, MobileViTv2 presents notable enhancements over its predecessor, MobileViTv1. MobileViTv2's adaptability together with its effective self-attention mechanism help to close the latency difference between conventional CNNs and ViT-based models on mobile devices while preserving competitive performance with less parameters. MobileViTv2 scales its network width using a width multiplier, $\alpha$, which ranges from 0.5 to 2.0, allowing for flexibility in model complexity and enabling adaptation to various hardware capabilities and performance requirements. This makes MobileViT a flexible and strong option for a broad spectrum of applications, from image classification to object recognition, especially in circumstances when computational resources are restricted. For our classification requirements, two particular variants—MobileViT 0.50 and MobileViT 0.75—offer customized solutions. These variants represent the width multipliers, $\alpha$, of 0.50 and 0.75, respectively. This scaling guarantees that the models keep a fair balance between computational efficiency and performance, so fitting for devices with different processing capability. It also has a robust performance compared to other ViT based models.

\paragraph{ResNet152}
Another pre-trained model we investigated in our study was ResNet152 containing 152 residual network blocks. This ResNet blocks provides a sophisticated neural network structure which can withstand the predicament of training deeper networks \citep{Mokni2022}. The residual block is composed of a sequence of convolutional layers, which are subsequently followed by batch normalization and ReLU activation functions. Residual learning is implemented to tackle the issue of the vanishing gradient. It has the ability to bypass connections that enable direct data transmission via the network without undergoing any changes at intermediate levels. The fundamental concept is to get knowledge of the residual, which is the discrepancy between the intended outcome and the present outcome, so facilitating the process of optimization \citep{Rahman2022,He2016,Podder2022}.

\subsubsection{Self-ONN}
\label{sec:self-onn}
To address the inherent limitations of CNN in terms of linearity, this novel approach, known as the Operational Neural Network (ONN)-based model, was introduced. The ONN is a hybrid network susceptible to learning intricate patterns from any given signal. It accomplishes this by leveraging a predefined collection of nonlinear operators. The ONN has shown enormous promise in many different fields, such as but not restricted to image denoising and restoration \citep{Kiranyaz2021}. Self-ONN constitutes a novel iteration of ONN \citep{Kiranyaz2021, Malik2021}. Compared to a predetermined set of operator archives, Self-ONN offers the potential to accumulate the optimal collection of operators across training. Therefore, the preceding results yield a framework that displays enhanced adaptability, enabling it to effectively navigate a broader spectrum of situations and demonstrate effective applicability in real-world contexts. Throughout the training phase, the operational layers ascertain the optimal selection of operators, which may consist of an amalgamation of traditional functions or unfamiliar functionalities. The expression for the output denoted as \( x_m^l \) at \( m \)-th neuron of \( l \)-th layer of any ONN can be expressed using the following equation:
\[
x_m^l = b_m^l + \sum_{i=1}^{N_{l-1}} \Psi_{mi}^l(w_{mi}^l, y_i^{(l-1)}) \tag{2}
\]
The biases and weights relating to a neuron and layer are symbolized by \( b_m^l \) and \( w_{mi}^l \), respectively. The input parameter of the prior layers has been designated by \( y_i^{(l-1)} \). The kernel size of the preceding layer can be expressed by \( N_{l-1} \). Nodal operator is denoted by \( \Psi_{mi}^l \). If the parameter \( \Psi_{mi}^l \) shows linearity in nature, the equation correlates to typical CNN. \( \Psi \) represents the nodal operator in the ONN framework, which can promptly be formulated through the following equation:
\[
\Psi(w, y) = w_1 \sin(w_2 y) + w_3 \exp(w_4 y) + \ldots + w_q y \tag{3}
\]
Where \( \mathbf{w} \) implies the \( q \)-dimensional array of factors, which consists of both internal parameters along with the weights associated with the different functions. The construction of \( \Psi \) can be achieved by adopting a Taylor series estimation of a function \( f(x) \), in the vicinity of a specific point \( x=b \):
\[
f(x) = f(b) + \frac{f'(b)}{1!}(x-b) + \frac{f''(b)}{2!}(x-b)^2 + \frac{f'''(b)}{3!}(x-b)^3 + \ldots \tag{4}
\]
Leveraging on Equation 4:
\[
\Psi(w, y) = w_0 + w_1 (y-b) + w_2 (y-b)^2 + \ldots + w_q (y-b)^q \tag{5}
\] 
Where, \( w_q = \frac{f^{(n)}(b)}{q!} \), represents the \( q \)-th parameter of the \( q \)-th order polynomial function. The Self-ONN model implements the tangent hyperbolic ($tanh$) as an activation function within an interval of [-1, 1] where a retains zero value.

\subsubsection{Self-DenseMobileNet}
\label{sec:selfdense}
\begin{figure}[H] % Use H to force placement at the exact location
    \centering
    % Subfigure 1 (Top)
    \begin{subfigure}[b]{1\textwidth}
        \centering
        \includegraphics[width=\textwidth]{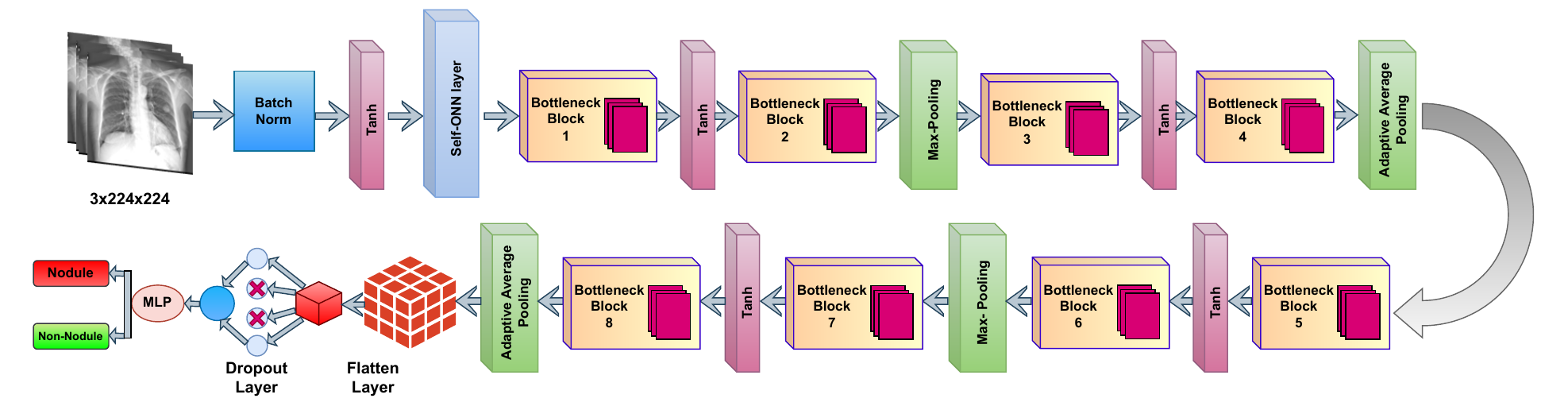}
        \caption{}
        \label{fig:self-densemobilenet-a}
    \end{subfigure}
    
    \vskip\baselineskip % Vertical space between subfigures
    
    % Subfigure 2 (Bottom)
    \begin{subfigure}[b]{1.1\textwidth}
        \centering
        \includegraphics[width=\textwidth]{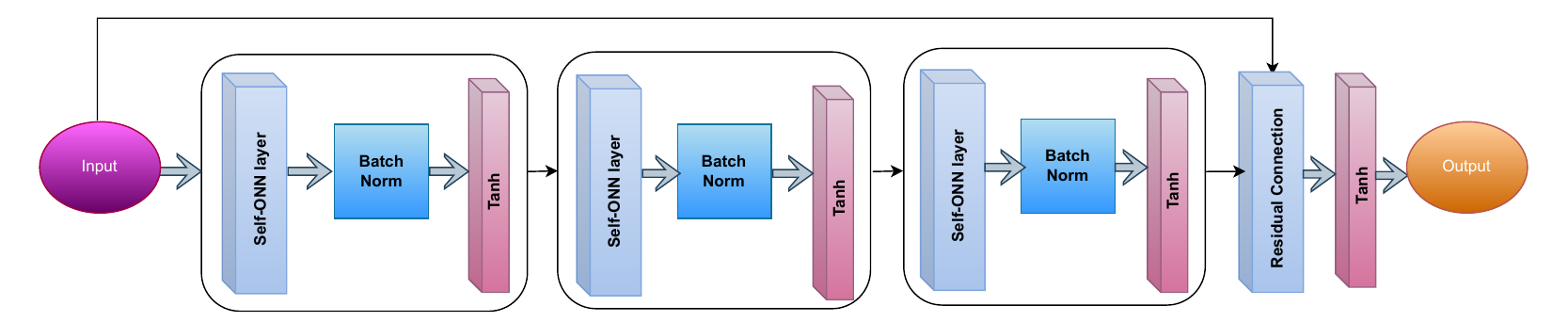}
        \caption{}
        \label{fig:self-densemobilenet-b}
    \end{subfigure}
    
    \caption{Self-DenseMobileNet Architecture and Bottleneck Block, (a) block diagram of the complete Self-DenseMobileNet architecture, (b) detailed view of the internal structure of the bottleneck block.}
    \label{fig:self-densemobilenet-main}
\end{figure}

We introduced the Self-DenseMobileNet architecture, which is built upon MobileNet by incorporating Self-ONN layers, along with a combination of bottleneck residual and non-residual blocks. Figure \ref{fig:self-densemobilenet-main} illustrates the block diagram of the proposed architecture. Self-DenseMobileNet is designed specifically for image classification tasks, incorporating adaptive average pooling and dropout for effective regularization. By integrating key features from Self-ONN, ResNet, and MobileNet, it offers a streamlined yet highly robust architecture. The model is shallow, meaning it reduces resource demands while still delivering high performance, making it both efficient and powerful for classification tasks.

% First Figure (Single image on top)

The architecture begins by processing images through Self-ONN layers, which utilize Self-ONN convolution, batch normalization, and tanh activation to extract important features. Bottleneck residual blocks, inspired by ResNet, refine these features by facilitating the reuse of information and improving gradient flow and optimization. Max pooling and adaptive average pooling layers reduce the spatial dimensions, preparing the data for classification. Unique features concatenation enables the model to capture patterns by combining characteristics from different layers. Regularization is achieved through Dropout, and the final classification is handled by a Self-MLP layer. While the model excels in efficiency and versatility across classification tasks, smaller models might trade off some accuracy compared to larger counterparts. Therefore, when choosing between efficiency and accuracy, specific use cases and resource constraints must be considered.

\subsubsection{Stacking Technique}
\label{sec:Classicalstacking}
\begin{figure}[!ht]
    \centering
    \includegraphics[width=.95\textwidth]{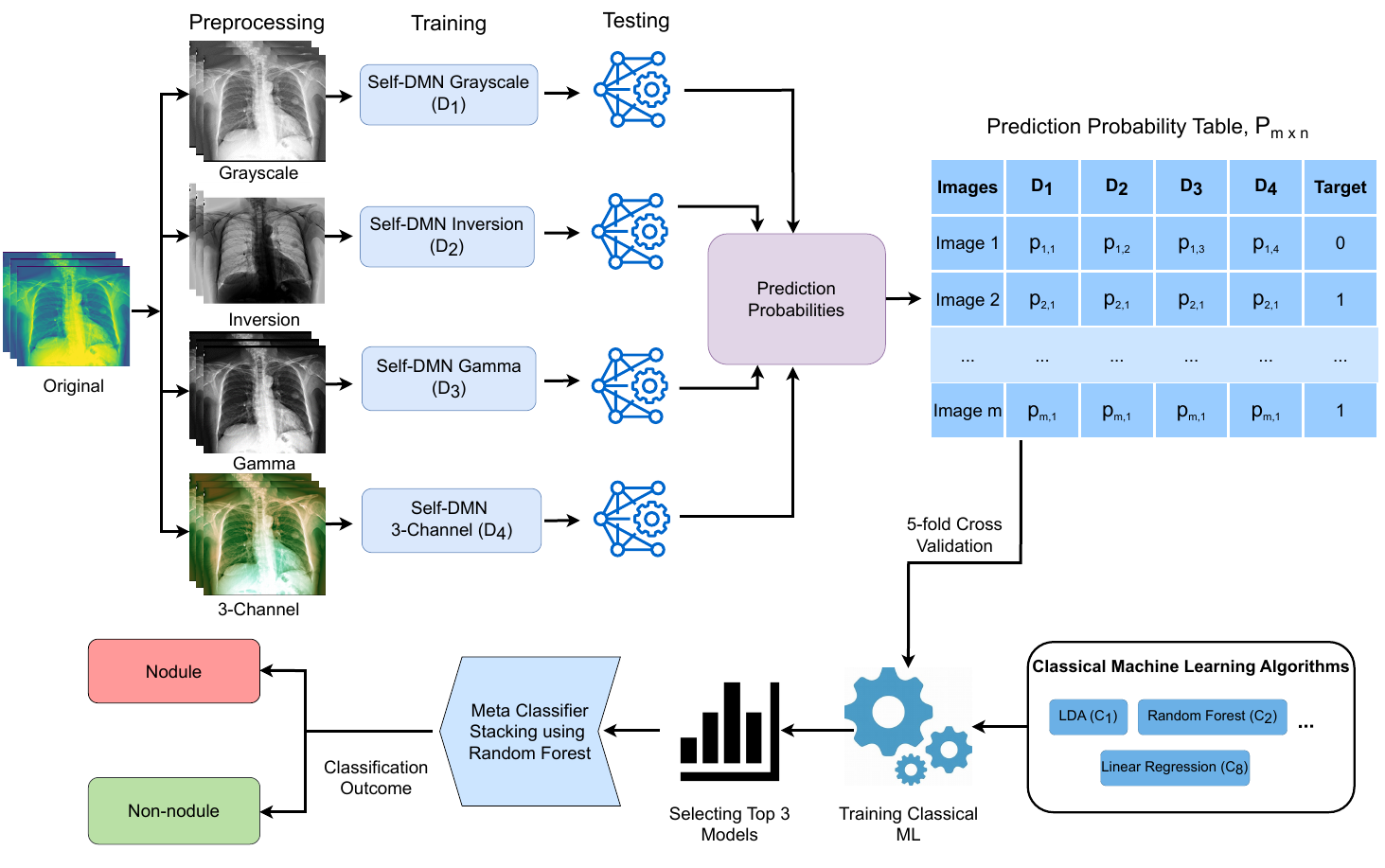}
    \caption{The process of stacking ensemble learning based nodule classification}
    \label{fig:stacking-architecture}
\end{figure}

To further enhance classification performance, we developed a meta-classifier using a stacking approach, as illustrated in Figure \ref{fig:stacking-architecture}. We trained four Self-DenseMobileNet models \( D_1, D_2, D_3, D_4 \) -- one for each variant of the images, Grayscale, Gamma corrected, Image colour inverted, and 3-Channel. These models are used to generate prediction probabilities \(P_{1,i}, P_{2,i}, P_{3,i}, P_{4,i} \) for an image $i$, forming a prediction probability table \( P_{m \times n} \), where \( m\) represents the number of images and \(n\) represents the number of columns on the table.
This dataset was used to train eight classical ML models \( C_1, C_2, \ldots, C_8 \) with 5-fold cross-validation. The eight classical ML models used in this study were Support Vector Machine (SVM)~\citep{Chen2015, gammermann2000support}, Multilayer Perceptron (MLP) \citep{Taud2018}, XGBoost \citep{Ren2017Jul}, Random Forest (RF) \citep{Li2023Feb}, Adaptive Boosting (Adaboost) \citep{schapire2013explaining}, Linear Discriminant Analysis (LDA) \citep{xanthopoulos2013linear}, Gradient Boosting \citep{Natekin2013}, and Logistic Regression (LR) \citep{Wright1995}.
%which are formulated in a table with a target column. 
The use of classical machine learning followed by a stacking algorithm was driven by their simplicity, interpretability and low computational cost, making them efficient for training the framework \citep{Li2022}. 
Based on their accuracy, the top three models were eventually used to build the meta-classifier using RF.
Stacking employs ensemble learning techniques, combining multiple models to enhance overall performance and deliver more robust and accurate results.
% A total of eight classical machine learning classifiers, such as  Support Vector Machine (SVM)\citep{Chen2015, gammermann2000support}, Multilayer Perceptron (MLP) \citep{Taud2018}, XGBoost \citep{Ren2017Jul}, Random Forest (RF) \citep{Li2023Feb}, Adaptive Boosting (Adaboost) \citep{schapire2013explaining}, Linear Discriminant Analysis (LDA) \citep{xanthopoulos2013linear}, Gradient Boosting \citep{Natekin2013}, and Logistic Regression (LR) \citep{Wright1995}, were utilized in this study. The three best-performing base models were selected based on their highest accuracies. These model prediction probabilities were then used as inputs for a meta-learner classifier. 
% Self-DenseMobileNet models \( D1, D2, D3, D4 \) trained on different image variants, provided the prediction probabilites \(P1, P2, P3, P4 \), forming a tabular dataset \( X_{m \times n} \), where \( m (=4)\) represents the number of different model outputs and \( n (=4)\) represents the number of images. This dataset was used to train all eight classical machine learning models \( C1, C2, \ldots, C8 \) with 5-fold stratified cross-validation. Based on their accuracy, the top three models were eventually used to build the meta-classifier \( M_f \).

\subsection{Model Interpretability}
\label{sec:cam}
Class Activation Mapping (CAM) provides a comprehensible visualization of how a deep learning model arrives at a specific decision during classification. This technique generates a heatmap on a given sample image, highlighting the areas that the trained model considers most important for identifying and classifying objects within the image. By pinpointing the specific regions influencing the model's decision, CAM enhances the transparency and interpretability of the classification process, allowing humans to better understand the reasoning behind the model's predictions.

Some of the most widely used CAM algorithms include SmoothGrad \citep{Smilkov2017}, Score-CAM \citep{Wang2020}, and Grad-CAM \citep{Selvaraju2017}. We selected Score-CAM due to its superior performance in classification problems involving medical images \citep{Fruh2021, Mullan2022}. Using Score-CAM generates a heatmap that visually highlights the areas of an image where the model has focused its learning. This explainablity approach enhances confidence in the model's decision-making by clearly demonstrating that it is learning from relevant regions within the image. This is a significant advantage over treating the model as a black box, where decisions are made without transparency.

\begin{figure}[ht]
    \centering
    \includegraphics[width=.8\textwidth]{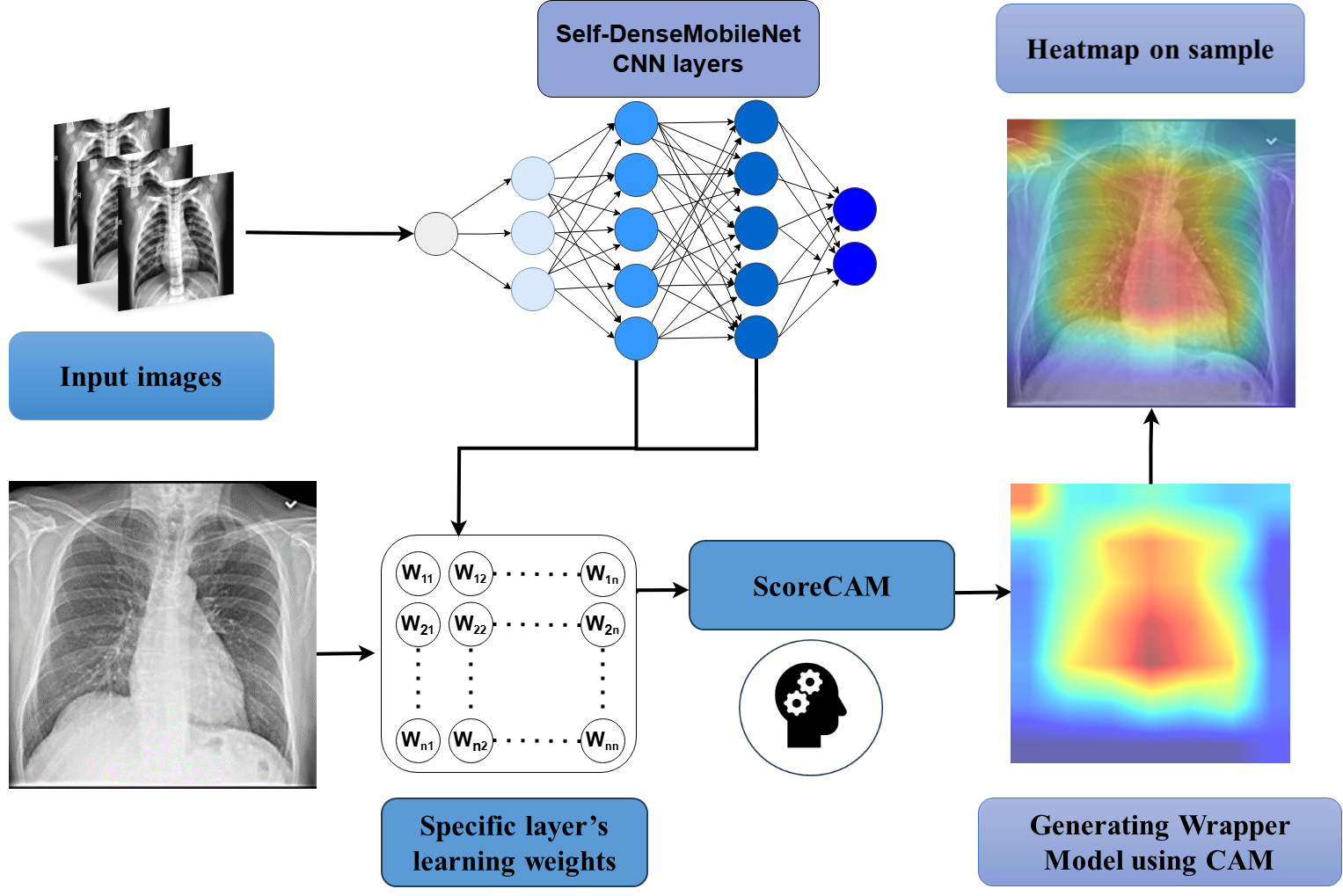}
    \caption{Interpretability of the deep learning model predictions.}
    \label{fig:Interpretability}
\end{figure}

Figure \ref{fig:Interpretability} illustrates the block diagram outlining the steps of the CAM algorithm. In our proposed framework, CXRs are passed to the Self-DenseMobileNet model. The model includes multiple hidden CNN layers that compute the network's weights and biases. These CNN layer weights are visualized through the ScoreCAM algorithm, which generates a heatmap that is superimposed on the original CXR image. The resulting image highlights key areas of interest, with red regions on the heatmap indicating the highest likelihood of a nodule.

\section{Experimental Evaluation}
\label{sec:experiment}

\subsection{Dataset split}
For a robust evaluation, we used 5-fold cross-validation. In each fold, the dataset was split with 795 nodule samples for training, 113 for validation, and 227 for testing, while the non-nodule samples consisted of 2,625 for training, 375 for validation, and 750 for testing.

\subsection{Environment and Training Parameters}
\label{sec:params}
The models were implemented using Python 3.10.12 and the PyTorch library. The computational resources utilized for this implementation included a system with 25 GB of RAM, a 12 GB T-4 GPU, and an Intel® Xeon® CPU operating at a frequency of 2.30 GHz. The training settings and termination criteria were consistent across all classification models, and each model underwent 50 epochs of back-propagation during the training process. Details of the training parameters are provided in Table \ref{tab:training_parameters}.
\begin{table}[!htb]
    \centering
    \caption{Training parameters for all the experiments}
    \label{tab:training_parameters}
    \begin{tabular}{ll}
    \toprule
    Parameters & Value \\
    \midrule
    Batch Size & 8 \\
    Learning Rate & 0.0001 \\
    Epochs & 50 \\
    Epochs Patience & 6 \\
    Stopping Criteria & 20 \\
    Optimizer & ADAM \\
    ImageNet & True \\
    \bottomrule
    \end{tabular}
\end{table}
\subsection{Performance Metrics}
\label{sec:evaluation}
% The performance of several classifiers was assessed using receiver operating characteristic (ROC) curves in conjunction with the area under the curve (AUC), as well as 
The performance of models was evaluated using metrics such as Precision, Sensitivity, Specificity, Accuracy, and F1-score. The results were based on the entire dataset, obtained through five-fold cross-validation, separated by Train, Validation and Test set. Given the class imbalance, weighted metrics for each class, as well as overall accuracy, were reported. 
% The area under the curve (AUC) was utilized as a statistic for performance comparison.
The mathematical formulations for the five evaluation metrics -- overall accuracy, precision, weighted sensitivity, F1 score and specificity  -- are presented in Equations 6-10 below:
\begin{align}
\text{Accuracy}_{\text{class } x} &= \frac{TP_{\text{class } x} + TN_{\text{class } x}}{TP_{\text{class } x} + TN_{\text{class } x} + FP_{\text{class } x} + FN_{\text{class } x}} \tag{6} \\
\text{Precision}_{\text{class } x} &= \frac{TP_{\text{class } x}}{TP_{\text{class } x} + FP_{\text{class } x}} \tag{7} \\
\text{Sensitivity/Recall}_{\text{class } x} &= \frac{TP_{\text{class } x}}{TP_{\text{class } x} + FN_{\text{class } x}} \tag{8} \\
\text{F1-score}_{\text{class } x} &= \frac{2 \times \text{Precision}_{\text{class } x} \times \text{Sensitivity}_{\text{class } x}}{\text{Precision}_{\text{class } x} + \text{Sensitivity}_{\text{class } x}} \tag{9} \\
\text{Specificity}_{\text{class } x} &= \frac{TN_{\text{class } x}}{TN_{\text{class } x} + FP_{\text{class } x}} \tag{10}
\end{align}
% \begin{flushleft}
% \footnotesize{where class \( x = \{\text{Nodule}, \text{Non-Nodule}\} \)}
% \end{flushleft}
Where, class $x$ is nodule or non-nodule. False Positive (FP), False Negative (FN), True Positive (TP), and True Negative (TN) refer to how well the model classifies instances as either belonging to the `nodule' (positive class) or `non-nodule' (negative class) categories. The positive class refers to the 'nodule' category, meaning instances where a lung nodule is present. TP represents cases where a nodule (positive class) is correctly identified, while TN refers to non-nodule instances (negative class) that are correctly classified. FP indicates instances where non-nodules are incorrectly classified as nodules, and FN indicates instances where nodules are incorrectly classified as non-nodules.

%Where, the False Positive (FP), False Negative (FN), True Positive (TP), and True Negative (TN) correspond to instances classified as positive or negative correctly or incorrectly. TP refers to instances from the positive class correctly identified, TN indicates instances from the negative class correctly identified, FP denotes instances from the negative class misclassified as positive, and FN reflects instances from the positive class misclassified as negative.%

% \section{Results}
% \label{sec:results}
% This section presents the results of multiple experiments conducted in the study, along with an analysis of the findings. The analysis includes the training of both deep learning models and classical models. Section \ref{sec:model_classification} provides the evaluation scores for the base DL models, comparing their performance with state-of-the-art models to highlight the iteration time of our novel proposed model. Section \ref{sec:stacking} details the stacking classification portion of the framework and compares our proposed framework with existing literature. Section \ref{sec:comparison} shows comparison of methodology and scores between several state-of-art literature. In Section \ref{sec:external}, we demonstrate the performance of our model on an external validation dataset. Finally, Section \ref{sec:cam_show} concludes by interpreting the results obtained from various models using CAM algorithms and displaying them on a heatmap.

\subsection{Self-DenseMobileNet Evaluation Results}
\label{sec:model_classification}
Table \ref{tab:model_performance} presents a comparison of the performance of the Self-DenseMobileNet model and several pre-trained CNN models for binary classification. The results that all the assessed pre-trained models demonstrate high levels of performance in accurately categorizing nodule and non-nodule in a binary-class scenario. Specifically, MobileViTv2-0.75 showed better scores in accuracy, precision, recall, and F1-score, only behind MobileViTv2-0.50 in specificity for grayscale samples. For image colour inversion samples, DenseNet201 outperformed all other models across all evaluation metrics. For gamma-corrected samples, ResNet152 outperformed MobileViTv2-0.50 in all models except for specificity. Similar trends were observed for 3-channel samples, where it led in all other evaluation metrics except specificity, where MobileViTv2-0.75 took the lead. Despite being trained from scratch, Self-DenseMobileNet exhibited comparable performance to transfer learning models in terms of evaluation criteria. The Self-DenseMobileNet model demonstrated impressive performance in handling inverted images, achieving an accuracy of 96.6\%, precision of 96.66\%, sensitivity of 96.6\%, F1-score of 96.62\%, and specificity of 95.04\%. Despite the Self-DenseMobileNet model's extremely simple design, these results were comparable to those of pre-trained models. We enhanced the performance of our proposed network with the incorporation of a stacking-based algorithm, as discussed in the following subsection.
\begin{table}[ht]
\centering
\caption{Performance metrics of different models under various image conditions.}
\label{tab:model_performance}
\resizebox{\textwidth}{!}{%
\begin{tabular}{llccccc}
\toprule
Image Condition & Metrics & DenseNet201 & MobileViTv2-0.50 & MobileViTv2-0.75 & ResNet152 & Self-DenseMobileNet \\ \midrule
\multirow{5}{*}{Grayscale} & Accuracy & 96.75 & 96.27 & \textbf{97.15} & 95.86 & 95.50 \\ 
% \cline{2-7} 
 & Precision & 96.72 & 96.50 & \textbf{97.15} & 95.84 & 95.68 \\ 
 % \cline{2-7} 
 & Sensitivity & 96.75 & 96.28 & \textbf{97.16} & 95.87 & 95.50 \\ 
 % \cline{2-7} 
 & F1-score & 96.73 & 96.33 & \textbf{97.15} & 95.85 & 95.55 \\ 
 % \cline{2-7} 
 & Specificity & 92.69 & \textbf{96.47} & 94.77 & 91.80 & 94.52 \\ \hline
\multirow{5}{*}{Inversion} & Accuracy & \textbf{97.54} & 96.66 & 96.95 & 96.66 & 96.60 \\ 
% \cline{2-7} 
 & Precision & \textbf{97.54} & 96.73 & 96.97 & 96.66 & 96.66 \\ 
 % \cline{2-7} 
 & Sensitivity & \textbf{97.55} & 96.66 & 96.95 & 96.66 & 96.60 \\ 
 % \cline{2-7} 
 & F1-score & \textbf{97.54} & 96.69 & 96.95 & 96.66 & 96.62 \\ 
 % \cline{2-7} 
 & Specificity & \textbf{95.51} & 95.25 & 95.14 & 93.77 & 95.04 \\ \hline
\multirow{5}{*}{Gamma} & Accuracy & 97.07 & 96.54 & 96.81 & \textbf{97.07} & 96.19 \\ 
% \cline{2-7} 
 & Precision & 97.06 & 96.68 & 96.90 & \textbf{97.08} & 96.21 \\ 
 % \cline{2-7} 
 & Sensitivity & 97.07 & 96.54 & 96.81 & \textbf{97.08} & 96.19 \\ 
 % \cline{2-7} 
 & F1-score & 97.06 & 96.58 & 96.83 & \textbf{97.06} & 96.20 \\ 
 % \cline{2-7} 
 & Specificity & 94.08 & \textbf{96.19} & 96.08 & 95.59 & 93.63 \\ 
 \hline
\multirow{5}{*}{3-Channel} & Accuracy & \textbf{96.99} & 95.91 & 96.97 & 95.60 & 96.34 \\ 
% \cline{2-7} 
 & Precision & 96.98 & 96.09 & \textbf{97.06} & 95.60 & 96.37 \\ 
 % \cline{2-7} 
 & Sensitivity & \textbf{96.99} & 95.91 & 96.97 & 95.60 & 96.34 \\ 
 % \cline{2-7} 
 & F1-score & 96.98 & 95.96 & \textbf{97.00} & 95.59 & 96.35 \\ 
 % \cline{2-7} 
 & Specificity & 94.11 & 95.32 & \textbf{96.32} & 95.37 & 94.16 \\ \bottomrule
\end{tabular}
}
\begin{flushleft}
\footnotesize
 *Bold values represent the best results attained among the trained models for a specific image condition in that evaluation metric \\
% *ACC = ACCURACY, PR = PRECISION, SENS = SENSITIVITY, F1 = F1-SCORE, SPE = SPECIFICITY \\
% *DNET201 = DENSENET-201, MVIT050 = MobileViTv2-0.50, MVIT075 = MobileViTv2-0.75, SELF-DMN = SELF-DENSEMOBILENET, RNET152 = RESNET152
\end{flushleft}
\end{table}

% Following the adoption of deep learning, two additional training sessions were integrated into the study. The predicted probabilities of Self-DenseMobileNet networks for different enhanced images were utilized as feature predictors for the nodule classification task during the classical ML training stage. Initially, we employed eight variations of conventional machine learning classifiers and subsequently constructed a novel stacking model. The top three models, chosen based on accuracy, were selected for the next training phase, and a novel stacking architecture was devised. The Random Forest (RF) classifier demonstrated superior performance in classifying nodules, achieving an accuracy of 98.65\%, precision of 99.73\%, recall of 98.51\%, sensitivity of 99.12\%, and an F1 score of 99.11\%. Next, the stacking model was built on top of the three best-performing models: RF, LDA, and LR. The stacking model achieved even better performance compared to the RF model, with an accuracy of 99.28\%, precision of 99.60\%, recall of 99.47\%, sensitivity of 98.68\%, and an F1 score of 99.53\%. 
% Table \ref{tab:model_performance} reports the efficacy of eight distinct state-of-the-art machine learning classifiers deploying stacking models for binary classification.

\begin{table}[ht]
\centering
\caption{Deep Learning model Inference Time analysis}
\label{tab:inference_time}
\begin{tabular}{lcc}
\toprule
Model & \shortstack{Inference time per image (ms) \\ time $\pm$ std} & \shortstack{Number of \\ trainable parameters} \\
\midrule
ResNet152 & 22.50 ± 6.71 & 58,147,906 \\
Self-DenseMobileNet & \textbf{10.15 ± 1.95} & 3,513,714 \\
DenseNet201 & 37.07 ± 37.78 & 18,096,770 \\
MobileViTv2-0.50 & 12.77 ± 2.01 & 1,114,107 \\
MobileViTv2-0.75 & 12.39 ± 2.18 & 2,481,779 \\
\bottomrule
\end{tabular}
\end{table}

The results in Table \ref{tab:inference_time} provide a comprehensive comparison of inference times and trainable parameters across several DL models. Self-DenseMobileNet emerges as the most efficient model, achieving the lowest inference time per image at 10.15 ms, significantly outperforming the other models in terms of speed. In particular, it shows more than a 2 times speed improvement over ResNet152 (22.50 ms) and nearly 4 times faster than DenseNet201 (37.07 ms), while also maintaining a much lower standard deviation, indicating more consistent performance. Despite having a slightly larger number of trainable parameters compared to lightweight models like MobileViTv2-0.50 (1,114,107 parameters) and MobileViTv2-0.75 (2,481,779 parameters), Self-DenseMobileNet's trade-off between trainable parameters and inference speed makes it highly suitable for practical deployment. In contrast, models like ResNet152 and DenseNet201, though more powerful in terms of trainable parameters (58,147,906 and 18,096,770 respectively), suffer from higher inference times and larger standard deviations. This can make them less viable for applications requiring fast, reliable predictions. The Self-DenseMobileNet model's ability to combine relatively low parameter count (3,513,714 parameters) with high efficiency makes it a versatile solution.

\subsection{Stacking Results}
\label{sec:stacking}
Following the adoption of DL, two additional training sessions were integrated into the study. The predicted probabilities of Self-DenseMobileNet networks for different enhanced images were utilized as feature predictors for the nodule classification task during the classical ML training stage. We employed eight variations of classical machine learning classifiers and subsequently constructed a novel stacking model. Table \ref{tab:stacking_performance} reports the efficacy of eight distinct state-of-the-art machine learning classifiers deploying stacking models for binary classification. The top three models, chosen based on accuracy, were selected for the next training phase, and a novel stacking architecture was devised. The Random Forest (RF) classifier demonstrated superior performance in classifying nodules, achieving an accuracy of 98.65\%, precision of 99.73\%, recall of 98.51\%, sensitivity of 99.12\%, and an F1 score of 99.11\%. Next, the stacking model was built on top of the three best-performing models: RF, LDA, and LR. The stacking model achieved even better performance compared to the RF model, with an accuracy of 99.28\%, precision of 99.60\%, recall of 99.47\%, sensitivity of 98.68\%, and an F1 score of 99.53\%. 

\begin{table}[ht]
\centering
\caption{Classical and Stacking models performance on Self-DenseMobileNet prediction probability CSV data.}
\label{tab:stacking_performance}
\begin{tabular}{lccccc}
\toprule
Classification model & Accuracy & Precision & Recall & Specificity & F1-Score \\
\midrule
Multilayer Perceptron Classifier (MLP) & 95.99 & 98.29 & 96.45 & 94.44 & 97.36 \\
Linear Discriminant Analysis (LDA) & 96.97 & 98.44 & 97.60 & 94.89 & 98.02 \\
XGBoost Classifier (XGB) & 96.21 & 98.66 & 96.37 & 95.68 & 97.50 \\
Random Forest (RF) & 98.65 & 99.73 & 98.51 & 99.12 & 99.11 \\
Logistic Regression (LR) & 96.62 & 98.62 & 96.96 & 95.50 & 97.78 \\
Support Vector Machine (SVM) & 62.66 & 89.50 & 58.19 & 77.43 & 70.53 \\
AdaBoost Classifier (ADA) & 96.27 & 98.77 & 96.34 & 96.03 & 97.54 \\
GradientBoosting Classifier (GB) & 96.25 & 98.06 & 97.04 & 93.65 & 97.55 \\
Meta-Random Forest Classifier (LDA+ RF + LR) & \textbf{99.28} & \textbf{99.60} & \textbf{99.47} & \textbf{98.68} & \textbf{99.53} \\
\bottomrule
\end{tabular}
\begin{flushleft}
\footnotesize

\end{flushleft}
\end{table}

\begin{figure}[!htb]
    \centering
    \includegraphics[width=0.6\textwidth]{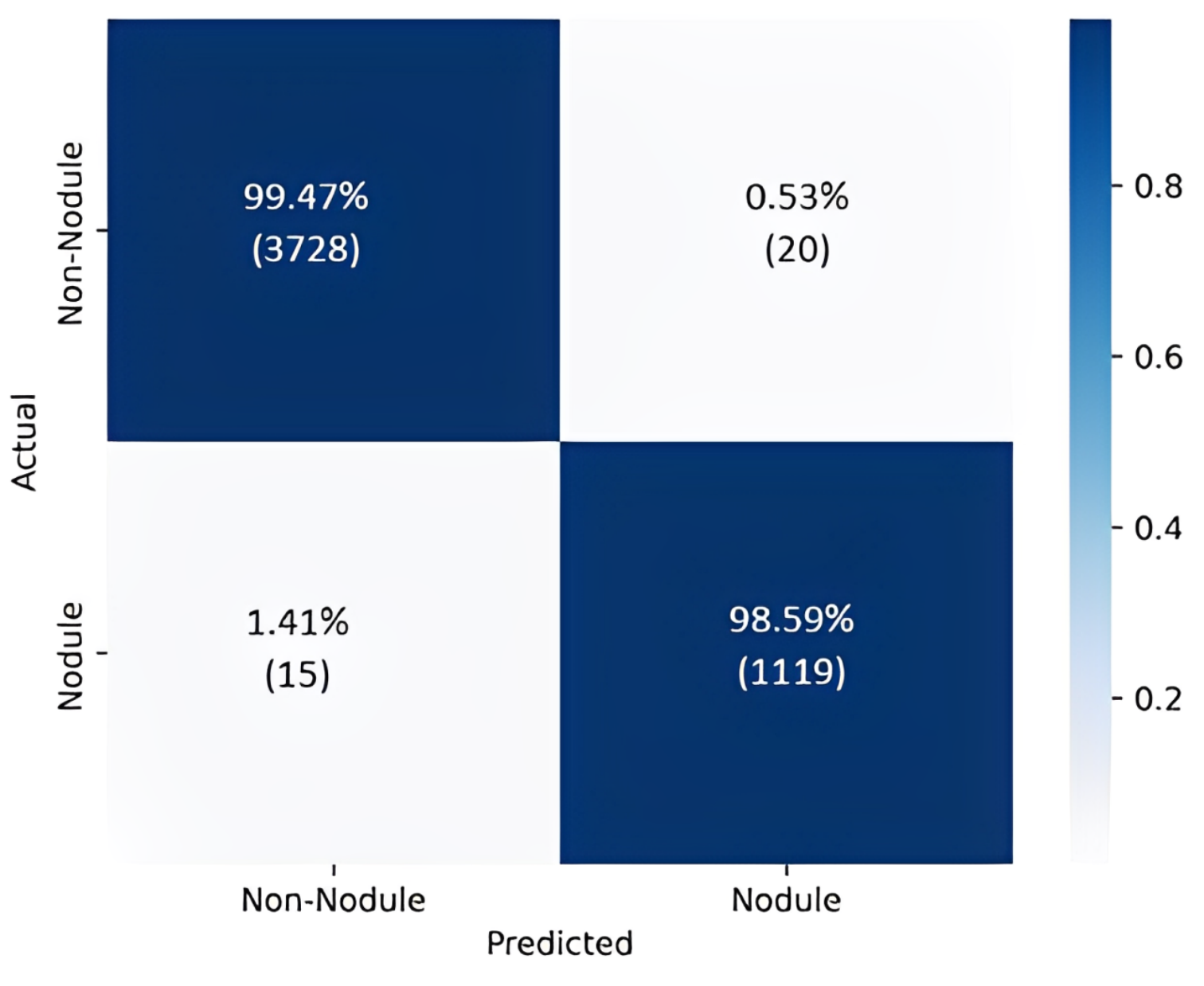}
    \caption{Confusion matrix displaying classification of nodule and non-nodule class for final Meta-Random Forest Classifier for internal validation}
    \label{fig:confusion_matrix}
\end{figure}

Figure \ref{fig:confusion_matrix} illustrates the confusion matrix of the stacking model, demonstrating its robust performance in classifying lung nodules and non-nodules. In this study, the confusion matrix illustrates the model's strong performance in classifying lung nodules and non-nodules. Following the standard layout, correct predictions for nodules, or TP, are found in the bottom-right, while accurate classifications of non-nodules, or TN, are located in the top-left. Misclassifications, though minimal, are reflected in the top-right, where FP non-nodules incorrectly identified as nodules—are recorded. Meanwhile, FN nodules misclassified as non-nodules—appear in the bottom-left. The model accurately predicted 1,119 nodule samples (TP) and 3,728 non-nodule samples (TN), with only 20 non-nodules misclassified as nodules (FP) and 15 nodules misclassified as non-nodules (FN). This resulted in an impressive true negative rate  of 99.47\% and a true positive rate  of 98.59\%. The False Positive Rate was as low as 0.53\%, and the False Negative Rate stood at 1.41\%, leading to an overall misclassification rate of just 1.94\%. These results reflect the model’s high reliability and precision in distinguishing between lung nodules and non-nodules, making it a robust tool for medical image analysis.
% Additionally, it depicts the ROC curve, revealing the AUROC values achieved by classical and stacking models. All models, except for MLP and SVM, obtained an AUROC of 0.99, showcasing the robustness of the proposed methodologies in distinguishing between the binary classes. 
% Notably, 99.47\% of the non-nodule samples and 98.59\% of the nodule cases were correctly classified, leaving little room for misclassification in internal validation, which only amounts to 1.94\% confusion.%

% \begin{figure}[ht]
% \centering
% \begin{subfigure}[b]{0.45\textwidth}
%     \centering
%     \includegraphics[width=\textwidth]{roc_curve.png}
%     \caption{ROC curve with their respective AUROC values for classical and stacking performance}
%     \label{fig:roc_curve}
% \end{subfigure}
% \quad
% \begin{subfigure}[b]{0.45\textwidth}
%     \centering
%     \includegraphics[width=\textwidth]{confusion_matrix.png}
%     \caption{Confusion matrix displaying classification of nodule and non-nodule class for final Meta-Random Forest Classifier for internal validation}
%     \label{fig:confusion_matrix}
% \end{subfigure}
% \caption{ROC curve and Confusion matrix}
% \label{fig:roc_confusion}
% \end{figure}

\subsection{External Validation} \label{sec:external}
We experimented with an external dataset to validate the authenticity of our framework's performance on completely unknown data. The external dataset, as discussed in Section \ref{sec:datasetdescription}, underwent identical preprocessing and enhancement approaches as used in internal validation. The Self-DenseMobileNet network, which exhibited comparable inference times to MobileViTv2-0.50 and MobileViTv2-0.75 (as observed in Table \ref{tab:external_validation}), was also evaluated in this external validation. Table \ref{tab:external_validation} represents the external validation performance of Self-DenseMobileNet, MobileViTv2-0.50, and MobileViTv2-0.75.

From Table \ref{tab:external_validation}, it is evident that Self-DenseMobileNet outperforms MobileViTv2-0.50 and MobileViTv2-0.75 across all types of image enhancements. Despite MobileViT models showing superior performance metrics on the trained dataset (as shown in Table \ref{tab:model_performance}), they demonstrate weaker capability on completely unseen datasets. For instance, Self-DenseMobileNet achieves 82.96\% accuracy, 84.62\% precision, 82.95\% sensitivity, 83.12\% F1-score, and 85.26\% specificity for Gamma-enhanced images, showcasing its potential in classifying external datasets. Additionally, our proposed stacking framework, leveraging the power of Self-DenseMobileNet, achieves even better scores with 89.40\% accuracy, 90.08\% precision, 92.58\% sensitivity, 91.31\% F1-score, and 84.6\% specificity.

\begin{table}[!ht]
\centering
\caption{Deep learning model's performance on an external dataset.}
\label{tab:external_validation}
% \resizebox{\textwidth}{!}{%
\begin{tabular}{llccc}
\hline
Image Condition & Metrics & MobileViTv2-0.50 & MobileViTv2-0.75 & Self-DenseMobileNet \\ \hline
Grayscale & Accuracy & 61.99 & 50.28 & \textbf{74.36} \\ 
% \cline{3-5} 
 & Precision & 67.10 & 50.61 & \textbf{80.48} \\ 
 % \cline{3-5} 
 & Sensitivity & 61.99 & 50.28 & \textbf{76.26} \\ 
 % \cline{3-5} 
 & F1-score & 50.53 & 50.43 & \textbf{74.28} \\ 
 % \cline{3-5} 
 & Specificity & 43.35 & 46.67 & \textbf{80.53} \\ \hline
Inversion & Accuracy & 68.05 & 58.96 & \textbf{83.98} \\ 
% \cline{3-5} 
 & Precision & 79.13 & 53.37 & \textbf{85.87} \\ 
 % \cline{3-5} 
 & Sensitivity & 68.05 & 58.97 & \textbf{83.99} \\ 
 % \cline{3-5} 
 & F1-score & 60.71 & 49.73 & \textbf{84.15} \\ 
 % \cline{3-5} 
 & Specificity & 51.75 & 42.02 & \textbf{86.69} \\ \hline
Gamma & Accuracy & 61.35 & 47.25 & \textbf{82.95} \\ 
% \cline{3-5} 
 & Precision & 59.83 & 50.81 & \textbf{84.62} \\ 
 % \cline{3-5} 
 & Sensitivity & 61.36 & 47.25 & \textbf{82.95} \\ 
 % \cline{3-5} 
 & F1-score & 59.70 & 47.48 & \textbf{83.12} \\ 
 % \cline{3-5} 
 & Specificity & 52.99 & 50.27 & \textbf{85.26} \\ \hline
3-Channel & Accuracy & 55.86 & 57.13 & \textbf{78.17} \\ 
% \cline{3-5} 
 & Precision & 54.38 & 54.80 & \textbf{82.43} \\ 
 % \cline{3-5} 
 & Sensitivity & 55.85 & 57.13 & \textbf{78.17} \\ 
 % \cline{3-5} 
 & F1-score & 54.77 & 54.95 & \textbf{78.33} \\ 
 % \cline{3-5} 
 & Specificity & 48.68 & 47.77 & \textbf{82.97} \\ \hline
\end{tabular}
% }
% \begin{flushleft}
% \footnotesize
% *IMAGE COND = IMAGE CONDITION. \\
% *ACC = ACCURACY, PR = PRECISION, SENS = SENSITIVITY, F1 = F1-SCORE, SPE = SPECIFICITY. \\
% *MVIT050 = MOBILEVIT-050, MVIT075 = MOBILEVIT-075, SELF-DMN = SELF-DENSEMOBILENET. \\
% *VALUES OF Q OF SELF-DMN: ORIGINAL = 7, INVERSION = 7, GAMMA = 5, 3-CHANNEL = 3.
% \end{flushleft}
\end{table}

% In Table \ref{tab:stacking_performance-externaldataset}, we selected the top 2 performing DL models from Table \ref{tab:model_performance} and incorporated our framework to compare with our proposed model's framework, where Self-DenseMobileNet dominates in terms of performance. This leads to the conclusion that despite the Vit-based models slightly outperforming our proposed model in terms of internal validation and number of parameters, they are outmatched in classifying completely unseen data compared to our proposed model.

With the intention to compare our newly proposed Self-DenseMobileNet architechture with existing DL models, we selected top 2 DL models from Table \ref{tab:model_performance}, MobileViTv2-0.75 and MobileViTv2-0.50 to evaluate our proposed stacking framework. Table \ref{tab:stacking_performance-externaldataset} clearly shows that Self-DenseMobileNet dominates the MobileVitv2 models in terms of performance. Meta-Random Forest Classifier (Self-DenseMobileNet) outshines the other 2 models by a large margin. This leads to the conclusion that despite the Vit-based models slightly outperforming our proposed framework in terms of internal validation and number of parameters, they are outmatched in classifying completely unseen data compared to our proposed framework. 

\begin{table}[!ht] % Use [H] to force the table to stay here
\centering
\caption{Stacking Classical model and final framework's performance on an external dataset.}
\label{tab:stacking_performance-externaldataset}
\begin{tabular}{llllll}
\toprule
Algorithm & Accuracy & Precision & Sensitivity & Specificity & F1-score \\ 
\midrule
Meta-Random Forest Classifier (Self-DenseMobileNet) & \textbf{89.40} & \textbf{90.08} & \textbf{92.58} & \textbf{84.6} & \textbf{91.31} \\ 
% \hline
Meta-Random Forest Classifier (MobileViTv2-0.75) & 52.51 & 54.63 & 52.51 & 52.88 & 53.00 \\ 
% \hline
Meta-Random Forest Classifier (MobileViTv2-0.50) & 70.23 & 69.90 & 70.23 & 66.32 & 69.99 \\ \bottomrule
\end{tabular}
\begin{flushleft}
\footnotesize
% * LDA = LINEARDISCRIMINANTANALYSIS, RF = RANDOM FOREST, LR = LOGISTIC REGRESSION. \\
% * MVIT050 = MobileViTv2-0.50, MVIT075 = MobileViTv2-0.75, SELF-DMN = SELF-DENSEMOBILENET.
\end{flushleft}
\end{table}

\begin{figure}[ht]
    \centering

    \includegraphics[width=0.6\textwidth]{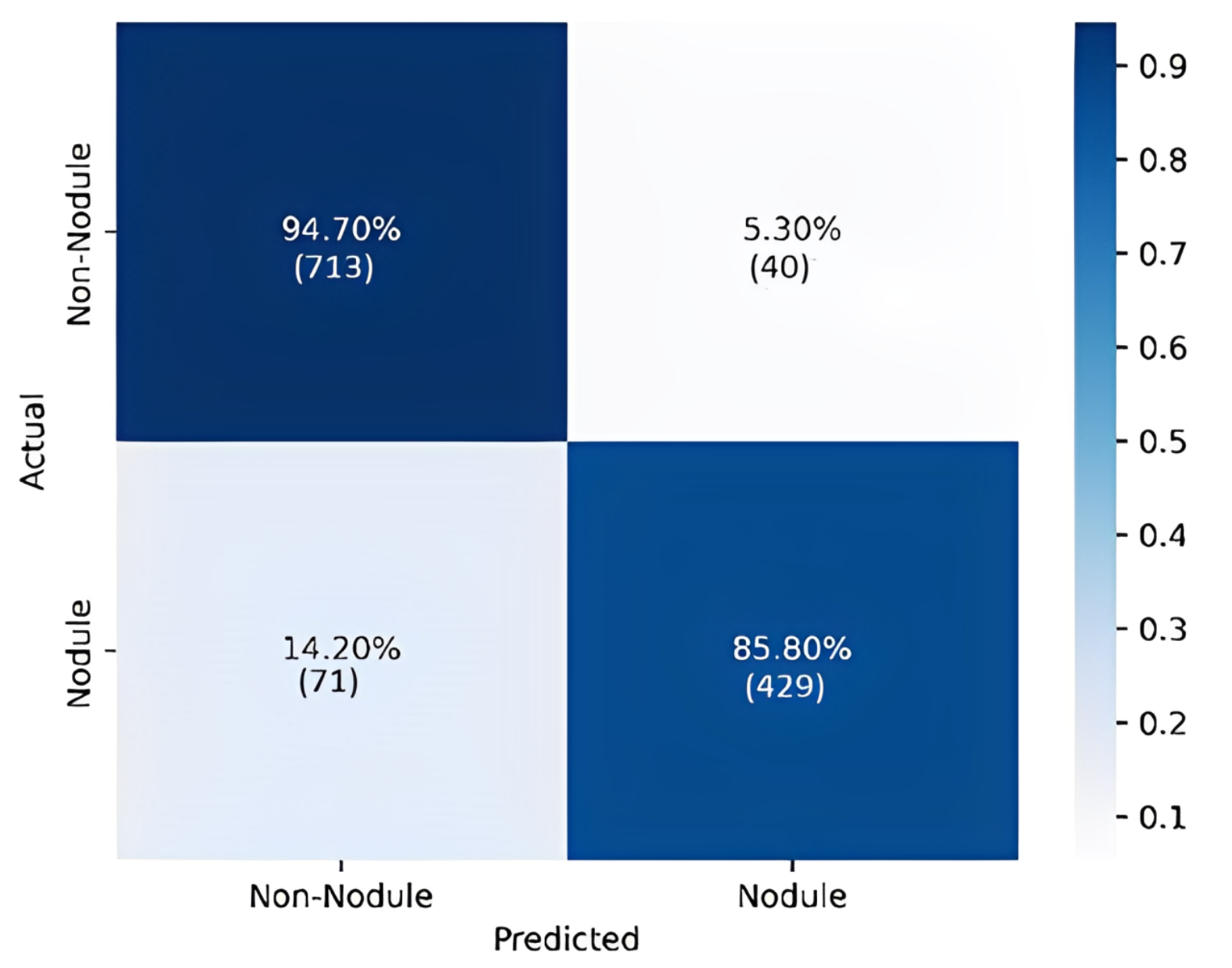}
    \caption{Confusion matrix for final proposed framework on external validation process.}
    \label{fig:confusion_matrix_self_dmn}
\end{figure}

The confusion matrix for the Meta-Random Forest Classifier for Self-DenseMobileNet, which combines classical ML models with a stacking algorithm, is shown in Figure \ref{fig:confusion_matrix_self_dmn}. The final proposed stacking model exhibits confusion in a cumulative 19.50\% of all external data.

\section{Discussion}
\subsection{Comparative Analysis Among Similar Works}
\label{sec:comparison}
Table \ref{tab:comparison} compares several notable works focusing on detecting and classifying lung nodules from CXRs and CT-scans. \citet{Ougul2015} studied the application of the LoG and SVM techniques to analyze chest radiographs from the JSRT dataset resulted in a sensitivity rate of 80\% while averaging 6.4 FPs, effectively addressing the challenge of nodule detection in radiographic images. A Fuzzy Hypersphere Neural Network (FHSNN) classifier was used to analyze chest radiographs within the JSRT dataset, demonstrating an accuracy rate of 83.35\%, showcasing the potential of FHSNN classifiers in medical image analysis \citep{Sonar2022}. CNNs were employed to evaluate chest radiographs from the JSRT dataset, resulting in an AUROC of 0.776, illustrating the strength of deep learning techniques in radiographic image analysis \citep{Mendoza2020}. Another approach using CNNs to analyze chest radiographs from the same dataset reported a notable accuracy rate of 86.67\%, further confirming the effectiveness of CNNs in enhancing the accuracy of radiographic images \citep{Thamilarasi2021}. A novel method by \citet{Monshi2021} introduced the CXR-labeler, a language model designed for medical image labeling, to analyze data from the MIMIC-CXR and PadChest datasets. This innovative approach achieved an impressive F1 score of 96.17\%, highlighting the potential of language models for medical image analysis. \citet{Wang2019} employed Thorax-net, a deep CNN, to assess chest radiographs within the ChestX-ray14 dataset. Their findings included an AUC of 0.7876 and 0.896, indicating the efficacy of deep CNNs in enhancing radiographic image analysis.

\begin{table}[ht]
\centering
\caption{Comparative analysis among similar works.}
\label{tab:comparison}
\begin{tabularx}{\textwidth}{@{} l X X @{}}
\toprule
Study & Approach & Results \\
\midrule
\citet{Schalekamp2014} & Commercial Software (ClearRead + Detect 5.2; Riverain Technologies) & 74\% sensitivity, AUC 0.841 \\
\citet{Schultheiss2020} & RetinaNet & ROC AUC value of 0.87 \\
\citet{Ougul2015} & Laplacian of the Gaussian (LoG), SVM & 80\% sensitivity with an average of 6.4 FPs \\
\citet{Sonar2022}& Fuzzy Hypersphere Neural Network (FHSNN) classifier & 83.35\% accuracy \\
\citet{Thamilarasi2021}& CNN & 86.67\% accuracy \\
\citet{Monshi2021} & CXR-labeler, a language model & 96.17\% F1 score \\
\citet{Wang2019} & Thorax-net, deep CNN & AUC of 0.7876 and 0.896 \\
\citet{Yan2019}& DenseNet, CNN & AUC 0.835 \\
\citet{Naik2020}& FractalNet & 94.7\% accuracy, 90.41\% specificity, 96.68\% sensitivity, AUROC 0.98 \\
\citet{Mary2024} & CNN, LSTM, Chaotic Population-based Beetle Swarm Algorithm & 0.9575 precision and 0.9646 sensitivity \\
\hline
This study & Image enhancement, augmentation, Self-DenseMobileNet, stacking-based meta-classifier & 99.28\% accuracy, 99.60\% precision, 99.47\% recall, 98.68\% specificity, and 99.53\% F1-score \\
\bottomrule
\end{tabularx}
\end{table}

\citet{Yan2019} utilized DenseNet to analyze radiographic data from the Picture Archiving Communication System of Tongji Affiliated Hospital and the PadChest dataset. This method demonstrated the capability of DenseNet in radiographic image interpretation, achieving an AUC of 0.835 . In another approach, FractalNet was applied to CT scans from LUNA dataset, achieving remarkable results. The model reached an accuracy of 94.7\%, specificity of 90.41\%, sensitivity of 96.68\%, and an AUROC of 0.98. These outcomes highlight the effectiveness of FractalNet in handling complex medical image analysis tasks \citep{Naik2020}. A different methodology combined CNNs, Long Short-Term Memory (LSTM) networks, and the Chaotic Population-based Beetle Swarm Algorithm to assess chest radiographs from the JSRT dataset. This integrated approach resulted in a precision of 0.9575 and a sensitivity of 0.9646, showing the potential of combining multiple techniques for better image analysis outcomes \citep{Mary2024}. Additionally, a CNN with deep feature fusion was employed to analyze CXRs from the JSRT dataset. The fusion-based approach reported a sensitivity of 69.2\% and a specificity of 96.02\%, demonstrating its effectiveness in improving radiographic image interpretation \citep{Wang2017}.

In contrast, our proposed CAD framework adopts a more comprehensive strategy by integrating image enhancement, augmentation, Self-DenseMobileNet, and a stacking-based meta-classifier to analyze radiographic data from the NODE21 Grand-challenge dataset. This framework achieved remarkable performance, with accuracy reaching 99.28\%, precision at 99.60\%, recall at 99.47\%, sensitivity at 98.68\%, and an F1-score of 99.53\%. These results demonstrate the robustness and potential of the framework in improving diagnostic accuracy in clinical settings.

\subsection{ScoreCAM Interpretation of Self-DenseMobileNet}
\label{sec:cam_show}
The calculative decision-making process of Self-DenseMobileNet has been facilitated through the application of the ScoreCAM visualizing method. ScoreCAM begins by calculating the importance or "score" of each class's prediction. This score quantifies how much the model relies on each class's evidence in the image to make its decision. A higher score indicates greater importance. The scoring is scaled into a feature map. ScoreCAM combines the feature maps with the class-specific scores by taking a weighted sum of the feature maps, where the weights are determined by the scores. Feature maps that are more important for the target class have a higher weight in the sum. The result of the weighted sum is an activation map, which highlights the regions in the image that contributed the most to the model's decision for the specific class. The regions highlighted in red show why the model concluded the image to contain a nodule. The heatmap serves as both a debugging tool and a human-interpretable method to identify nodules in the lung region. Upon inspecting the images, we can clearly depict the result outcomes.
\begin{figure}[!ht]
    \centering
    % Insert your ScoreCAM heatmap images here
    \begin{subfigure}[b]{0.45\textwidth}
        \centering
        \includegraphics[width=\textwidth]{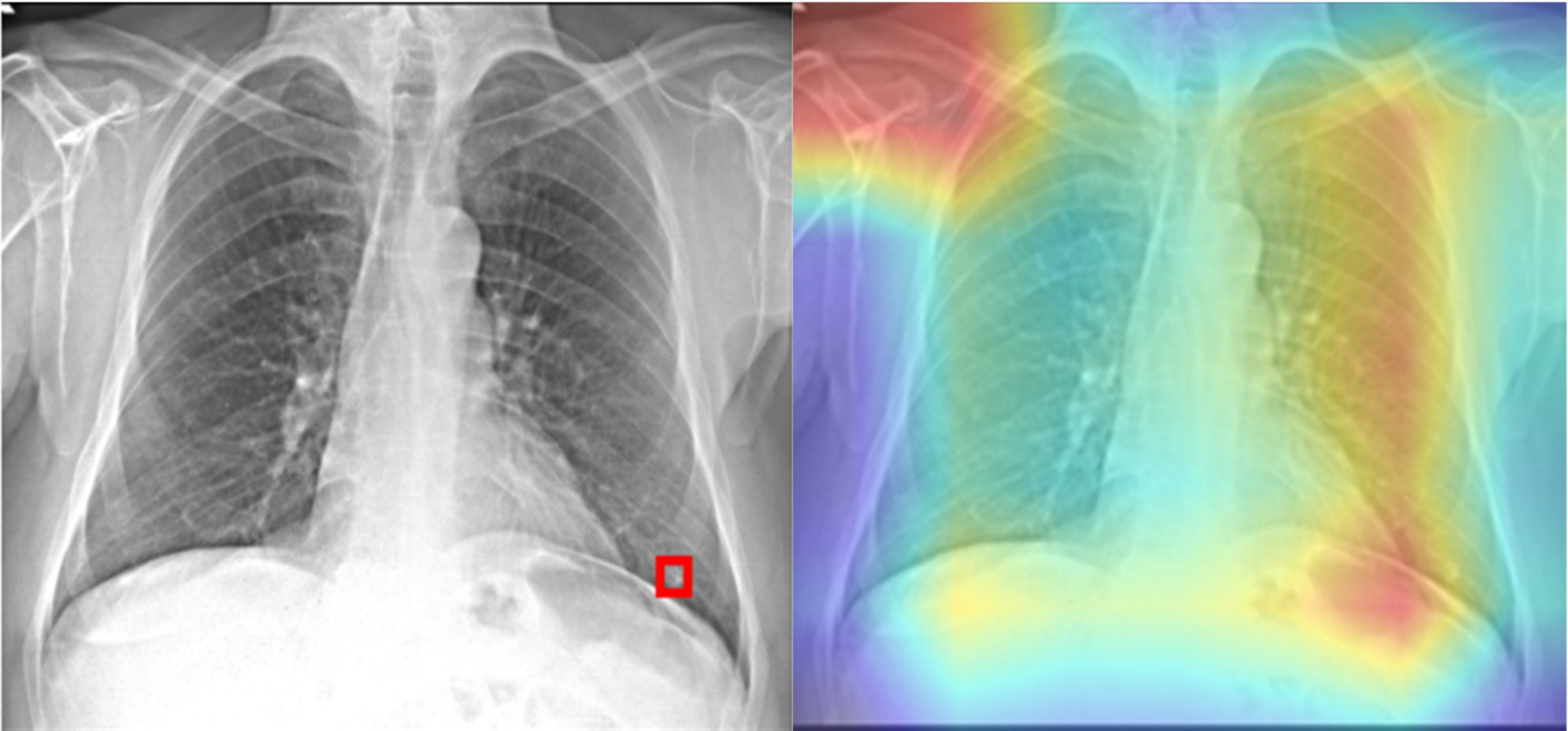} % Replace with your image path
        \caption{}
        \label{fig:cam1}
    \end{subfigure}
    \begin{subfigure}[b]{0.45\textwidth}
        \centering
        \includegraphics[width=\textwidth]{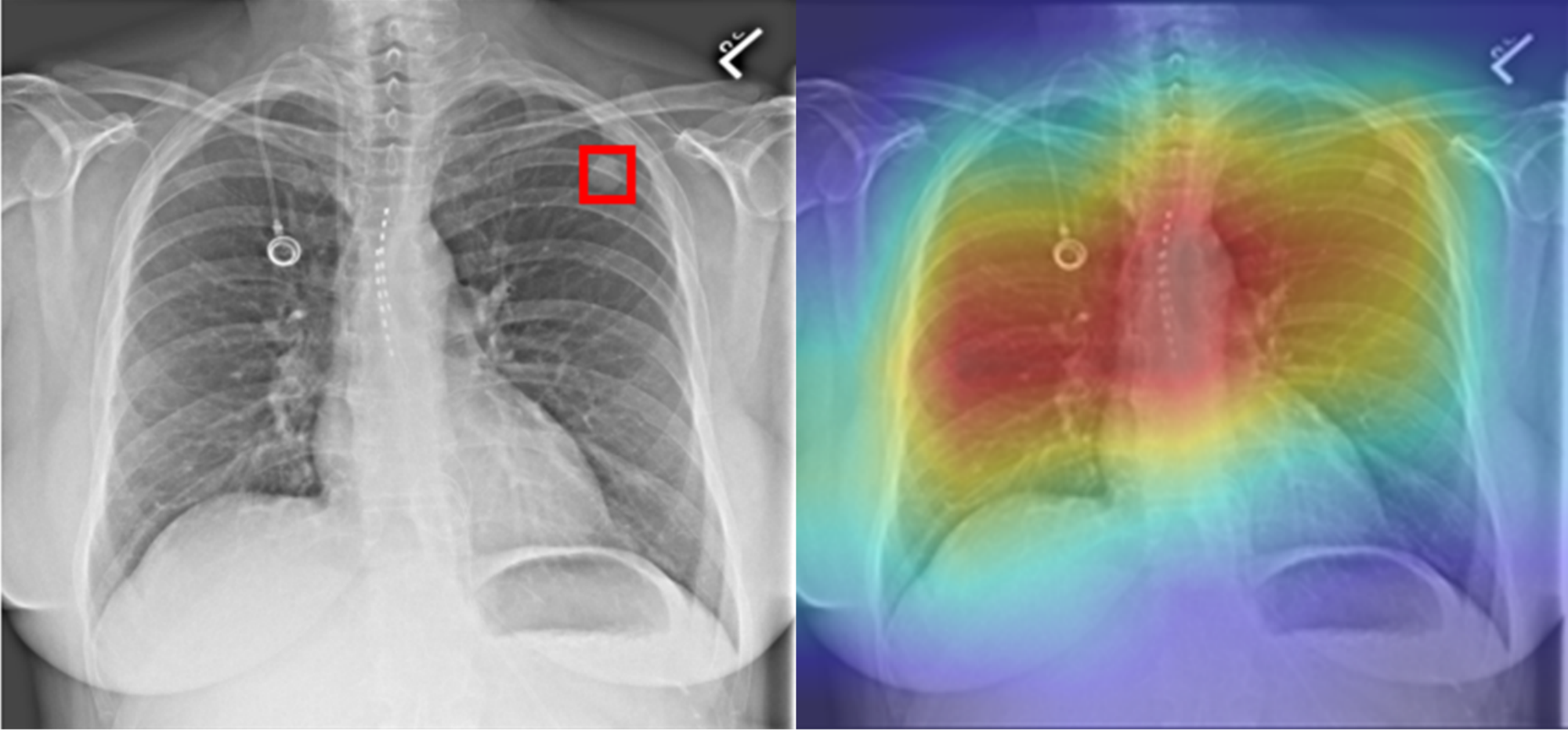} % Replace with your image path
        \caption{}
        \label{fig:cam2}
    \end{subfigure}
    \begin{subfigure}[b]{0.45\textwidth}
        \centering
        \includegraphics[width=\textwidth]{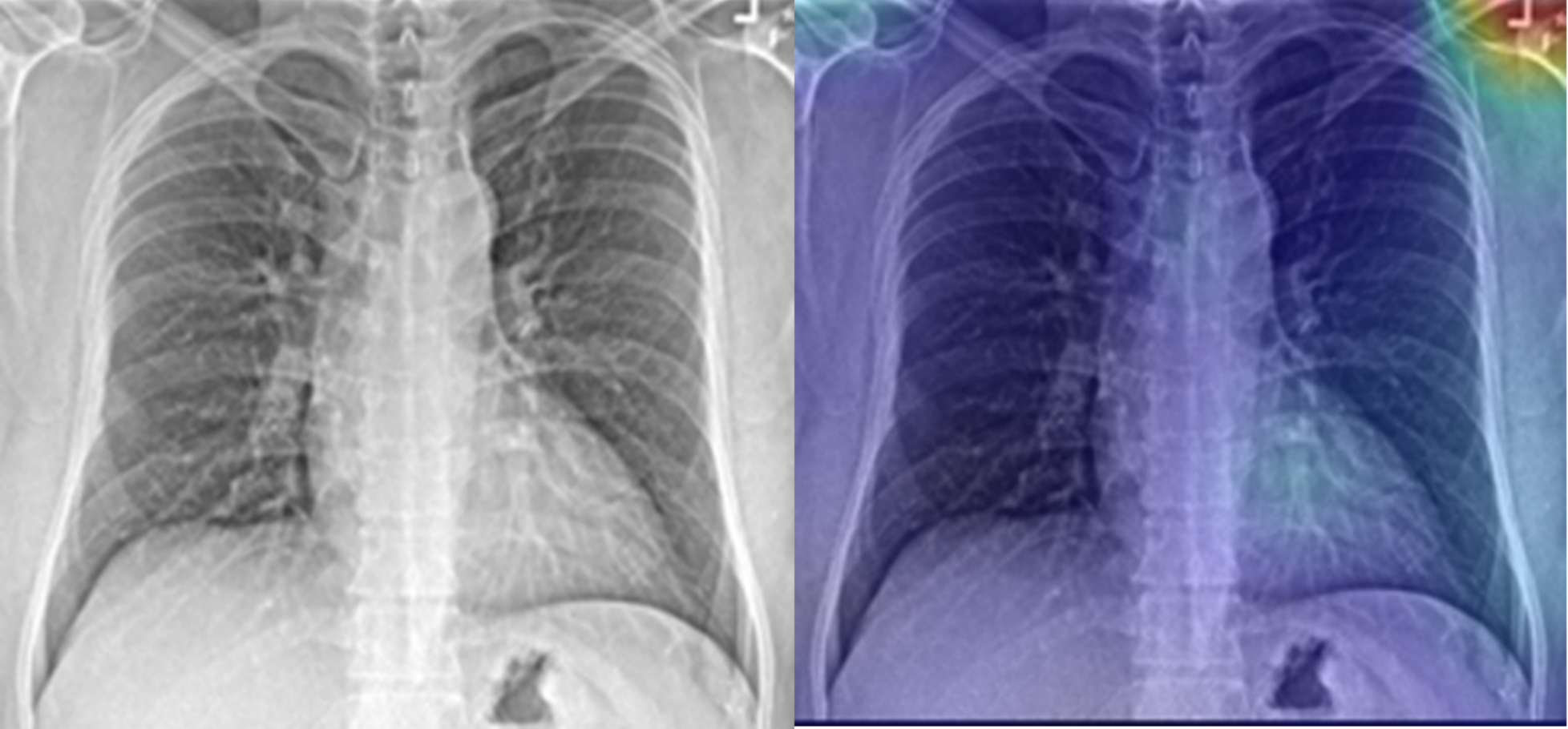} % Replace with your image path
        \caption{}
        \label{fig:cam3}
    \end{subfigure}
    \begin{subfigure}[b]{0.45\textwidth}
        \centering
        \includegraphics[width=\textwidth]{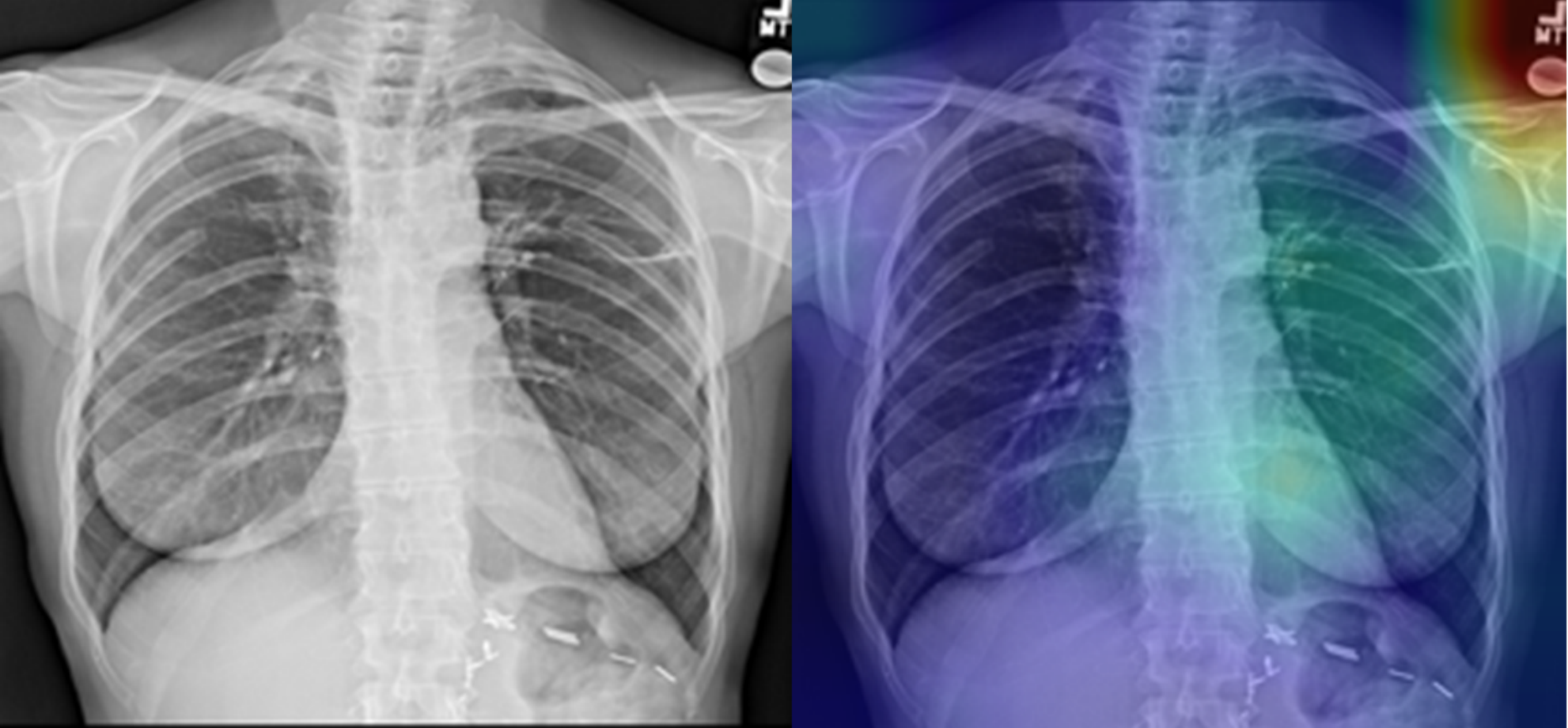} % Replace with your image path
        \caption{}
        \label{fig:cam4}
    \end{subfigure}
    \caption{Heatmaps generated by ScoreCAM. (a) shows a nodule on the bottom right lung, marked in red. (b) depicts a nodule on the upper right lung, with the heatmap marking a foreign object around the rib. (c) and (d) show no nodules, as indicated by the heatmap.}
    \label{fig:scorecam_heatmaps}
\end{figure}

Figure \ref{fig:scorecam_heatmaps} shows the ScoreCAM visualization of selected images in our dataset. It demonstrates how the model identifies areas of anomaly and scans for nodules. If it identifies a nodule within the red-marked region, the image is classified as “Nodule” class. In cases where foreign objects like pacemakers or vascular devices are present in CXRs, the model initially highlights them but later focuses solely on the nodule region. This highlights the robustness of our proposed model.

\section{Conclusion}
\label{sec:conclusion}
Addressing the pressing issue of accurately identifying radiographs containing cancerous nodules is imperative. The proliferation of CAD frameworks offers a promising solution to enhance radiographic interpretation. While a substantial body of research has concentrated on nodule detection, the prevalence of false positives remains a persistent challenge, often arising from variations in data quality. To combat this issue, we embarked on devising an optimal approach for isolating radiographs featuring cancerous nodules with both high specificity and accuracy, ultimately mitigating the problem of false positives within CAD frameworks.

In our study, we curated a diverse database compilation to broaden the diversity of our image dataset. Since radiographs are typically grayscale images, we initially focused on enhancing image quality. Our exploration of various medical imaging processes enlightened us on the intricacies of establishing a novel, multi-combination channel for image quality enhancement. Additionally, we approached this challenge by implementing deep learning methods, which proved instrumental in conducting a rigorous investigation. We conducted experiments on an external dataset to validate our framework, yielding consistent performance metrics. Moreover, employing CAM methods enabled us to identify distinct regions within the images that significantly influenced the models' decisions. Collectively, our framework demonstrates its reliability in the majority of chest radiographs.

In conclusion, classifying CXRs for Lung Nodules is indispensable in aiding radiologists in their diagnostic pursuits. Our proposed framework will significantly enhance patient outcomes, healthcare efficiency, and the future development of accurate nodule classification systems.

\section*{Data Availability Statement}
The dataset used in this study is publicly available at \url{https://node21.grand-challenge.org/Data/}.

\section*{Funding}
This research did not receive any specific grant from funding agencies in the public, commercial, or not-for-profit sectors.

\section*{Declaration of competing interest}
The authors declare no conflict of interest.

% \section*{Ethics Declaration}
% This article is based on previously conducted studies and does not contain any studies with human participants or animals. The dataset used in this research is collected from several open access and restricted access datasets (upon request) where standard protocols were used for creating the dataset.
\section*{CRediT authorship contribution statement}
\textbf{Md. Sohanur Rahman:} Conceptualization, Methodology, Data Curation, Software, Formal Analysis, Writing - Original Draft Preparation, Writing - Reviewing and Editing.
\textbf{Muhammad E. H. Chowdhury:} Conceptualization, Methodology, Supervision, Writing - Original Draft Preparation, Writing - Reviewing and Editing, Funding Acquisition
\textbf{Hasib Ryan Rahman:} Conceptualization, Methodology, Software, Writing - Original Draft Preparation, Writing - Reviewing and Editing.
\textbf{Mosabber Uddin Ahmed:} Conceptualization, Methodology, Supervision, Writing - Reviewing and Editing.
\textbf{Muhammad Ashad Kabir:} Conceptualization, Methodology, Supervision, Validation, Supervision, Writing - Reviewing and Editing.
\textbf{Sanjiban Sekhar Roy:} Validation, Writing - Reviewing and Editing.
\textbf{Rusab Sarmun:} Conceptualization, Methodology, Software, Writing - Original Draft Preparation, Writing - Reviewing and Editing.
%All authors have read and approved the final manuscript.

% \bibliographystyle{elsarticle-num-names} 
% \bibliographystyle{elsarticle-num}
\bibliographystyle{elsarticle-harv}
\bibliography{reference}
\end{document}